\documentclass[twocolumn, aps, prb, superscriptaddress]{revtex4-2}

\usepackage{subfigure}
\usepackage{microtype}
\usepackage{physics2}
\usephysicsmodule{ab}
\usephysicsmodule{ab.braket}
\usepackage{comment}
\usepackage{esvect}
\usepackage{siunitx}
\usepackage[version=4]{mhchem}
\usepackage[usenames, dvipsnames]{color}
\usepackage{epsfig}
\usepackage{graphicx}
\usepackage{float} 
\usepackage{gensymb}
\usepackage{multirow}
\usepackage{dcolumn}   
\usepackage{bm}        
\usepackage{ulem}
\usepackage[colorlinks=true,citecolor=blue,linkcolor=blue,urlcolor=blue]{hyperref}
\newcommand*{\vb}[1]{\boldsymbol{#1}} 

\begin{document}

\title{Phase-dependent electronic structure of two-dimensional Ag layers at the graphene/SiC interface}

\author{Sawani Datta}
\email{s.datta@fkf.mpg.de}
\affiliation{Max-Planck-Institut für Festkörperforschung, Heisenbergstra\ss e 1, 70569 Stuttgart, Germany}

\author{Boyang Zheng}
\affiliation{Department of Physics, The Pennsylvania State University, University Park, PA 16802, USA}
\affiliation{2-Dimensional Crystal Consortium, The Pennsylvania State University, University Park, PA 16802, USA}

\author{Arpit Jain}
\affiliation{Department of Materials Science and Engineering, The Pennsylvania State University, University Park, PA 16802, USA}

\author{Kathrin K\"uster}
\affiliation{Max-Planck-Institut für Festkörperforschung, Heisenbergstra\ss e 1, 70569 Stuttgart, Germany}

\author{Joshua A. Robinson}  
\affiliation{2-Dimensional Crystal Consortium, The Pennsylvania State University, University Park, PA 16802, USA}
\affiliation{Department of Materials Science and Engineering, The Pennsylvania State University, University Park, PA 16802, USA}
\affiliation{Center for Atomically Thin Multifunctional Coatings, The Pennsylvania State University, University Park, PA 16802, USA}

\author{Vincent H. Crespi}
\affiliation{Department of Physics, The Pennsylvania State University, University Park, PA 16802, USA}
\affiliation{2-Dimensional Crystal Consortium, The Pennsylvania State University, University Park, PA 16802, USA}
\affiliation{Department of Materials Science and Engineering, The Pennsylvania State University, University Park, PA 16802, USA}

\author{Ulrich Starke}
\affiliation{Max-Planck-Institut für Festkörperforschung, Heisenbergstra\ss e 1, 70569 Stuttgart, Germany}

\begin{abstract}
Intercalation at the graphene/SiC interface provides a controlled route to stabilize atomically thin layers with properties distinct from their bulk counterparts. 
In this platform, the structure and stability of the intercalated phase depend sensitively on the defect landscape of the starting substrate. 
For intercalated two-dimensional silver at the graphene/SiC interface, two phases have been observed: a phase epitaxial to the SiC lattice, called Ag$_{(1)}$, readily obtained following the conventional intercalation method under ultra-high-vacuum conditions and extensively characterized, and a more densely packed phase, called Ag$_{(2)}$, which has remained largely unexplored.
Here we report an \textit{in situ} ultra-high-vacuum preparation method of the second phase intercalated at the graphene/SiC interface; this phase previously was prepared via high-pressure confinement heteroepitaxy. 
Low-energy electron diffraction shows that Ag$_{(2)}$ is rotated by \ang{30} relative to the SiC lattice and forms supercells, in contrast to the $(1\times 1)$ epitaxial relation of Ag$_{(1)}$ with SiC. 
High-resolution angle-resolved photoemission spectroscopy reveals a more complex Ag$_{(2)}$ band dispersion compared to the Ag$_{(1)}$, as a consequence of the modified Ag-Si interaction in the denser phase. 
In density functional theory calculations, by defining the unfolding entropy which, in a quantified way, finds that the band structure of Ag$_{(2)}$ is more suitable to be unfolded to the SiC primitive cell, and the resulting unfolded band dispersion is in great agreement with the experimental data.
We further show that the different intercalated Ag phases tune the electronic properties of the overlying quasi-free-standing graphene layer differently: compared with Ag$_{(1)}$, Ag$_{(2)}$ yields an $\sim$1.75 times higher charge carrier density and modifies the charge-plasmon interaction of the graphene layer, indicating a change in effective screening at the interface.
\end{abstract}

\maketitle

\section{Introduction}

Quantum confinement in reduced-dimensional systems is a central topic in materials science, driven by both fundamental questions \cite{Novoselov2004, Geim2007, Son2006, Bhimanapati2015, Lin2016} and technological prospects \cite{Bera2010, Huang2022, Novoselov2005, karakachian_one-dimensional_2020}. 
In particular, reducing the film thickness to the atomic limit constrains electron motion and reshapes the electronic density of states, which can markedly modify chemical reactivity, mechanical response, and optical properties relative to the bulk \cite{Parzinger2017, Lee2008, Mak2010}. 
From a fundamental perspective, atomically thin layers can host emergent phenomena absent in their three-dimensional counterparts, including altered screening, enhanced many-body interactions, and symmetry-driven electronic instabilities \cite{Bostwick2010, Rosenzweig2022, Link2019, Herrera2024}. 
A key challenge, however, is achieving and maintaining the structural and chemical stability of such two-dimensional (2D) phases, which are often metastable \cite{Zhong2017, Romanyuk2009, Koerner2011} and strongly influenced by the choice of substrate, interface bonding, and environmental exposure \cite{Aizawa2014, Riedl2009}. 
In this context, intercalating a two-dimensional material at the graphene/SiC interface offers a practical way to improve stability. 
The SiC substrate can help stabilize an intercalated layer by providing interfacial bonding and an ordered template for growth, while the graphene overlayer acts as a largely inert, atomically thin cap that shields the intercalated film from the environment and lowers its chemical reactivity \cite{ Riedl2009, Emtsev2011, Forti2020, Rosenzweig2020, Matta2022, albalushi2016, El-Sherif2021, Vera2024, Wundrack2026, Lu2026, Pompei2025}.

As noted above, substrate templating--and thus the quality and defect landscape of the starting substrate--plays a central role in stabilizing intercalated interfacial layers. 
In this context, early studies showed that 2D-Ag and 2D-Ga intercalated at the graphene/SiC interface using the confinement heteroepitaxy (CHet) approach (performed under elevated pressure) \cite{Briggs2020} can form two distinct monolayer (ML) phases \cite{Wetherington2021, Zhang2025}. 
Subsequent systematic work on CHet-grown Ag-MLs demonstrated that the occurrence of these phases depends extensively on the type and density of defects in the initial substrate \cite{Jain2025}. 
In particular, substrates dominated by line and boundary defects tend to yield an Ag-ML, with a $(1\times 1)$ epitaxial relationship to the SiC lattice, which we refer to as the Ag$_{(1)}$ phase. 
In contrast, increasing the density of $sp^3$-type defects favors formation of a denser Ag-ML \cite{Jain2025}. In this Ag$_{(2)}$ phase, Ag adopts a more complex relationship with the SiC unit cell and is rotated by \ang{30} with respect to SiC \cite{Jain2025}. 
These two phases of Ag also exhibit significant differences in the optical responses, visible electronic absorption and sensing applications \cite{Zhang2025, Liu2025, Jain2025}. 
Combined experimental and theoretical studies further indicate that Ag$_{(2)}$ is thermodynamically more stable, and a time-dependent transformation from Ag$_{(1)}$ to Ag$_{(2)}$ has been reported \cite{Jain2025}.

While the CHet method enables the formation of intercalated layers over large areas at the graphene/SiC interface \cite{Briggs2020, Dong2024}, the present work focuses on an \textit{in situ} ultra-high vacuum (UHV) route that is directly compatible with stepwise preparation and immediate surface-sensitive characterization by LEED, ARPES, and XPS without breaking vacuum. 
In this sense, CHet and UHV intercalation provide complementary routes: CHet is well suited for scalable synthesis, whereas UHV enables the \textit{in situ} structural and electronic property measurements to avoid any undesired phase transitions in the ambient condition \cite{Jain2025}. 
Using standard UHV intercalation (Ag deposition on zero-layer graphene followed by annealing), the Ag$_{(1)}$ phase is typically obtained \cite{Rosenzweig2020,Rosenzweig2022}. 
To our knowledge, however, preparation of Ag$_{(2)}$ entirely under UHV conditions has not been demonstrated so far.
Here we present the first UHV preparation route that yields Ag$_{(2)}$ intercalation at the graphene/SiC interface, enabled by deliberately increasing vacancy-type defects through an additional intercalation step as described in the main text. 
We then provide a detailed experimental and theoretical investigation of the structural and electronic properties of atomically thin 2D-Ag$_{(2)}$. 
The structural characteristics are determined by low-energy electron diffraction (LEED) and supported by density functional theory (DFT) calculations. 
The electronic properties are investigated using high-resolution angle-resolved photoemission spectroscopy (ARPES) in combination with DFT band-structure calculations. 
The calculated band structure based on the proposed  structural model for Ag$_{(2)}$ \cite{Jain2025} reproduces the key features of the ARPES spectra.
As expected, Ag$_{(2)}$ exhibits a more complex electronic structure than Ag$_{(1)}$, accompanied by a clear shift in the Ag core-level spectrum. 
We further show that the phase-selective intercalation of Ag at the graphene/SiC interface provides a route to tune the $n$-type doping of quasi-free-standing monolayer graphene (QFMLG) and modifies the electron-plasmon coupling in graphene via proximity effects. 
Finally, we demonstrate that supercell formation between Ag$_{(2)}$ and graphene, produces replica features in ARPES, consistent with superstructure signatures observed in the LEED.

\section{Methods}
\subsection{Sample preparation}
For the preparation of graphene, we start with a mechanically polished and $n$-doped 6H-SiC (0001) substrate (from SiCrystal GmbH) with a miscut of $< 0.1$$\degree$ with respect to the $\{0001\}$ plane.
Following a chemical HF cleaning, to achieve an atomically flat surface for homogeneous graphene growth, the polishing scratches were removed by annealing the sample in an 800 mbar hydrogen atmosphere at about \qty{1405}{\degreeCelsius} for 20 minutes \cite{Ramachandran1998, soubatch2005}. 
The graphitisation of the SiC sample was performed in the same reactor under an Ar atmosphere (800 mbar pressure) by sublimating the Si atoms \cite{Emtsev2009} through heating the sample to \qty{\sim1445}{\degreeCelsius} for 4.5 minutes. 
This technique results in the formation of a graphene buffer layer (zero layer graphene or ZLG) with terraces a few microns wide \cite{Riedl2010, Forti2014}. 
The ZLG exhibits the ($6\sqrt{3}\times 6\sqrt{3}$)$R\ang{30}$ reconstruction relative to the underlying SiC substrate originating from the lattice mismatch \cite{Riedl2010, Emtsev2008}. Due to the presence of the partial $sp^3$ bonding at the ZLG/SiC interface, ZLG would not show any Dirac cone dispersion in the ARPES data \cite{Riedl2010, Emtsev2008}.

For Ag$_{(1)}$ intercalation, Ag was deposited for \qty{30}{min} (nominal rate \qty{1}{\angstrom/min}) onto a ZLG sample held at \qty{400}{\degreeCelsius}, followed by post-annealing at \qty{600}{\degreeCelsius} and \qty{650}{\degreeCelsius} for \qty{30}{min} at each temperature. 
A second deposition step with the same Ag dose was then carried out at \qty{400}{\degreeCelsius}, followed by annealing from \qty{600}{\degreeCelsius} to \qty{700}{\degreeCelsius} in \qty{50}{\degreeCelsius} steps, holding for \qty{30}{min} at each temperature.
This two-step preparation protocol follows the previously reported refined procedure \cite{Rosenzweig2022} for getting sharper Ag and graphene bands. 
Although the annealing temperatures and times were kept essentially the same, the present samples were prepared in a different preparation chamber with a modified experimental setup. 
In particular, we used electron-beam heating instead of direct plate heating as the reported one, and the evaporator-to-sample distance also differed. 
Nevertheless, the characteristic LEED signatures and the electronic band structures of both the Ag-ML and the resulting QFMLG are well reproduced. 
We observe a slight shift of the Ag-derived bands to higher BE compared to ref.\cite{Rosenzweig2022}.

For Ag$_{(2)}$, Ag was first evaporated for \qty{20}{min} (nominal rate \qty{1}{\angstrom/min}) onto the ZLG sample held at \qty{450}{\degreeCelsius}, followed by sequential annealing at \qty{600}{\degreeCelsius}, \qty{630}{\degreeCelsius}, and \qty{650}{\degreeCelsius} for \qty{20}{min} at each temperature.
Up to this point, the procedure is essentially identical to the one-step procedure used in ref.\cite{Rosenzweig2020} for Ag$_{(1)}$ intercalation, and the LEED pattern likewise shows the characteristic Ag$_{(1)}$-QFMLG signature: enhanced graphene (1$\times$1) spots and suppressed ($6\sqrt{3}\times 6\sqrt{3}$)$R\ang{30}$ features (see Supplementary Information (SI), section A, Fig. S1 (a) and (b)). 
As mentioned above, to obtain the Ag$_{(2)}$ phase in UHV, the defect landscape of the initial substrate is modified, in particular by promoting additional vacancy-type defects \cite{Briggs2020, Liu2021}. 
To achieve this, we introduced one cycle of Ag deintercalation followed by re-intercalation. 
Deintercalation was carried out using Pb, taking advantage of the fact that the heavier Pb atoms can replace intercalated Ag at the interface. 
Specifically, Pb was deposited onto the Ag$_{(1)}$-QFMLG sample at room temperature for \qty{5}{min} (nominal rate \qty{4}{\angstrom/min}), followed by annealing at \qty{400}{\degreeCelsius} and \qty{500}{\degreeCelsius} for \qty{30}{min} at each temperature~\cite{Matta2022}.
These steps result in complete substitution of Ag by Pb, yielding Pb-QFMLG (cf. SI, section A, Fig. S1 (c)), in agreement with the established LEED and ARPES signatures reported previously \cite{Matta2022, Matta2025}. 
In the subsequent Ag intercalation step, Ag was deposited onto the Pb-QFMLG sample at \qty{450}{\degreeCelsius} for \qty{30}{min} (nominal rate \qty{1}{\angstrom/min}), followed by annealing at \qty{600}{\degreeCelsius}, \qty{650}{\degreeCelsius}, and \qty{700}{\degreeCelsius} for \qty{30}{min} at each temperature.
Notably, in separate preparations we found that omitting the initial Ag-intercalation step (i.e., starting directly from Pb-QFMLG) still yields the same final Ag$_{(2)}$ phase; the first Ag-intercalation cycle therefore appears to mainly create additional defect/entry channels that facilitate the subsequent Pb intercalation.

Thus, the key practical difference between Ag$_{(1)}$ and Ag$_{(2)}$ in UHV is the starting substrate: Ag$_{(1)}$ is obtained by intercalating Ag into ZLG, whereas Ag$_{(2)}$ is obtained by intercalating Ag into Pb-QFMLG. 
Qualitatively, this can be understood by noting that a higher density of vacancy-type defects in Pb-QFMLG, from the process of Pb intercalation, likely provides favorable entry channels and nucleation sites for stabilizing Ag$_{(2)}$ intercalation \cite{Matta2025, Fiori2017}. 
In addition, because Pb forms a supercell with graphene \cite{Schaedlich2023,Matta2022, Vera2024, Matta2025}, associated  boundary defects could further promote \cite{Chen2020} the formation of the Ag$_{(2)}$ phase. 
A quantitative characterization of the defect density in Pb-QFMLG is beyond the scope of the present work. 
Future studies could address this point and explore whether other intercalated QFMLG systems can serve as alternative starting substrates for stabilizing the Ag$_{(2)}$ phase.

\subsection{Characterization}
We investigated the initial sample quality and analyzed the structural information \textit{in situ} by LEED using an Er-LEED system (SPECS GmbH).
The experimental photoemission spectroscopy (PES) data of the Ag$_{(2)}$ intercalated sample presented in this work were acquired at the Bloch beamline of the MAX IV synchrotron facility (Lund, Sweden) \cite{Polley2024}. 
On the Ag$_{(1))}$ intercalated sample, PES was measured at the $1^2$ end-station of UE112 PGM-2a-$1^2$ beamline at BESSY II, Helmholtz-Zentrum Berlin, Germany \cite{Varykhalov2018}. 
To prevent air exposure during transport, the samples were transferred in a Ferrovac UHV suitcase with a base pressure below \qty{1e-9}{mbar}.
All PES measurements were carried out at a sample temperature \qty{\approx 20}{K}. 
At the Bloch beamline, PES measurements were performed using a DA30-L hemispherical analyzer (Scienta Omicron GmbH). 
Synchrotron radiation with photon energies in the range \qtyrange{40}{180}{eV} was used for ARPES, while photon energies of \qtyrange{180}{650}{eV} were employed for core-level x-ray photoemission spectroscopy (XPS). 
The best overall energy resolution achieved in ARPES was approximately \qty{18}{meV}. 
At the $1^2$ end-station, ARPES measurements were done using an R8000 hemispherical analyzer (Scienta Omicron GmbH). 
Measurements were performed with photon energies of \qty{110}{eV} and \qty{40}{eV}, and the best overall energy resolution achieved was \qty{20}{meV}.   

\subsection{DFT calculation}

The calculations are done by the Vienna \textit{ab initio} simulation package (VASP)~\cite{kresseEfficiencyAbinitioTotal1996,kresseEfficientIterativeSchemes1996,kresseInitioMolecularDynamics1993,kresseUltrasoftPseudopotentialsProjector1999}, using the projector augmented wave (PAW) method~\cite{blochlProjectorAugmentedwaveMethod1994} and the Perdew-Burke-Ernzerhof (PBE) exchange-correlation functional~\cite{perdewGeneralizedGradientApproximation1996}.
The computational unit cells have 3 layers of SiC stacked in the ABC-form (consistent with the 6H-SiC) whose lattice constant in the $ab$-plane is \qty{3.096}{\angstrom}, and the $z$-dimension of the cell is \qty{20}{\angstrom} to include enough vacuum preventing interactions between the periodic images.
The bottom layer is passivated by H atoms and no graphene cap is included.
For Ag$_{(1)}$, we used the ideal $(1\times1)$ registry of Ag on top of the hollow-site of the topmost SiC layer.
For Ag$_{(2)}$, we have ($3\sqrt{3}\times3\sqrt{3}$) (in total 27) Ag atoms on top of a $(5\times5)$ SiC supercell.
The cutoff energy is \qty{450}{eV}, the self-consistent field (SCF) convergence criterion is \qty{1.4e-8}{eV\per atom}, and the force convergence criterion is \qty{0.01}{eV/\angstrom} with D3 van der Waals correction~\cite{grimmeConsistentAccurateInitio2010,grimmeEffectDampingFunction2011}.
The k-point samplings are $\Gamma$-centered \numproduct{13x13x1} (\numproduct{51x51x1}) for the self-consistent (non-self-consistent) calculations for Ag$_{(1)}$, and $\Gamma$-centered \numproduct{3x3x1} (\numproduct{10x10x1}) for Ag$_{(2)}$.
The band unfolding~\cite{popescuExtractingEffectiveBand2012} is done with the \texttt{VaspBandUnfolding} code~\cite{zhengVaspBandUnfolding2023}.
Since we don't have the graphene cap, or take into account of any charge transfer effect, the extra electrons from Ag in Ag$_{(2)}$ will be in the conduction band, while experimentally the Fermi level is in the gap.
In addition, our calculations do not include any correction for the underestimated band gap problem in DFT~\cite{perdewDensityFunctionalTheory1985}.
Therefore, we ignore the conduction bands when comparing with ARPES.

\section{Results and discussion}
\subsection{Structural and chemical characterization of Ag$_{(2)}$ intercalated QFMLG}
\begin{figure*}[htbp]
	\vspace{-0.5 cm}
	\centering
	\includegraphics[width=1.05\textwidth]{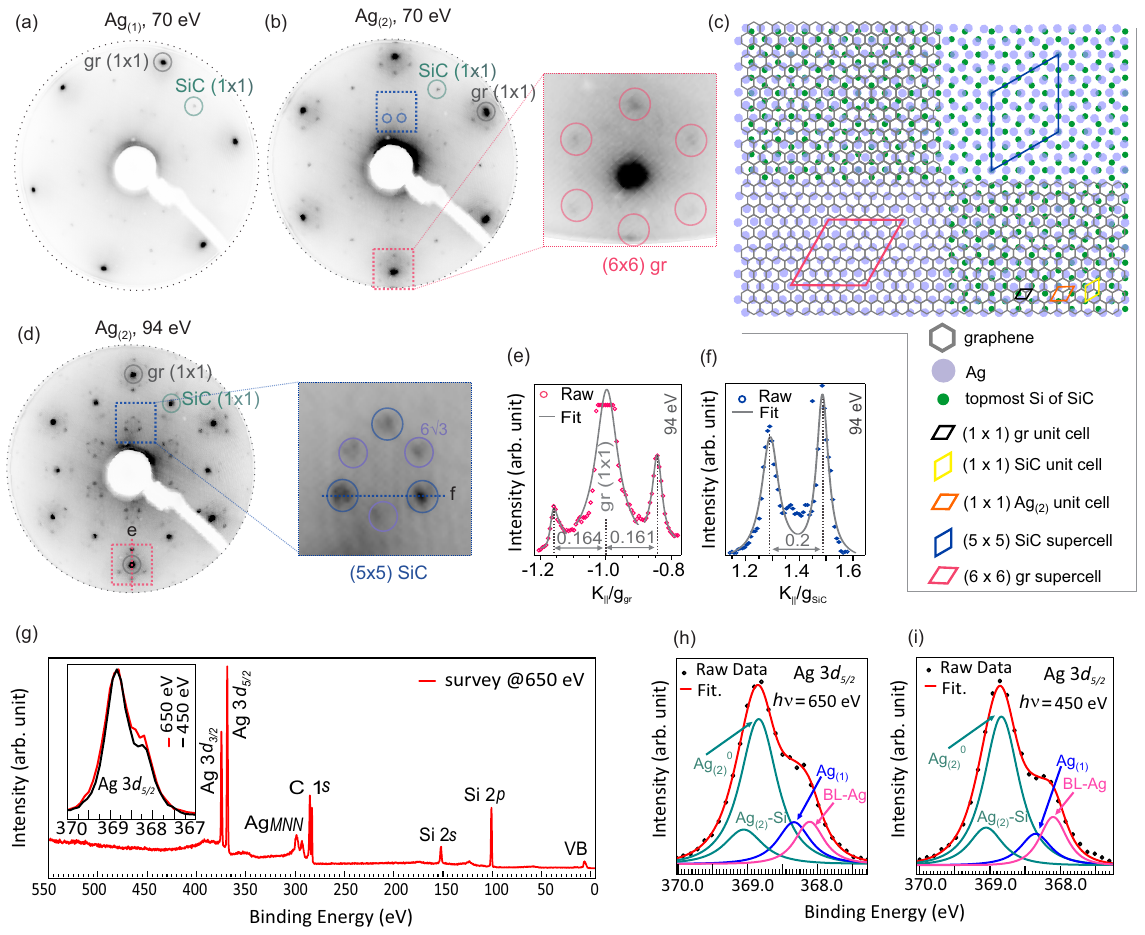}
	\vspace{-12.8 cm}
	\caption{
	LEED patterns of (a) Ag$_{(1)}$-QFMLG and (b) Ag$_{(2)}$-QFMLG at \qty{70}{eV}. 
	The graphene $(1\times 1)$ and SiC $(1\times 1)$ spots are indicated by gray and green circles, respectively, in (a), (b), and (d). The $(6\times 6)$ graphene replica spots are highlighted by pink circles in the enlarged view associated with (b), while the $(5\times 5)$ SiC replica spots are marked by blue circles.
	(c) Top-view schematic of the SiC/Ag$_{(2)}$/graphene stacking sequence. 
	Upper left: complete stack, showing the Si atoms (green spheres) of the topmost SiC, Ag$_{(2)}$ atoms (violet spheres), and the graphene lattice (gray hexagon). 
	Upper right: Ag$_{(2)}$ on top of Si(C), including the $(5\times 5)$ SiC supercell (blue rhombus). 
	Lower left: graphene on Ag$_{(2)}$ with the $(6\times 6)$ graphene supercell (pink rhombus). 
	Lower right: combined view including all unit cells; the legend is shown at the bottom.
	(d) LEED pattern of Ag$_{(2)}$ at \qty{94}{eV}; the $(5\times 5)$ SiC replica spots are marked in the corresponding enlarged view.
	(e, f) Line profiles (raw data: pink/blue open circles; fit: gray solid lines) taken through the (e) $(6\times 6)$ graphene replicas and (f) $(5\times 5)$ SiC replicas, along the directions indicated in (d). 
	(g) XPS survey spectrum of Ag$_{(2)}$ acquired at \qty{650}{eV}; inset: Ag~3$d_{5/2}$ region measured at \qty{650}{eV} (red) and \qty{450}{eV} (black). 
	(h, i) Fits to the Ag$_{(2)}$ 3$d_{5/2}$ spectra at (h) \qty{650}{eV} and (i) \qty{450}{eV}, with the corresponding fit components shown. }
	\label{fig1}
\end{figure*}

We begin with a structural comparison of the two intercalated 2D-Ag phases using LEED. 
Figures~\ref{fig1}(a) and \ref{fig1}(b) show LEED patterns of the Ag$_{(1)}$- and Ag$_{(2)}$-intercalated samples, respectively, recorded at the surface-sensitive electron energy of \qty{70}{eV} \cite{Seah1979}. 
For Ag$_{(1)}$ (Fig.~\ref{fig1}(a)), the pattern is dominated by graphene $(1\times 1)$ reflections (grey circle), while the SiC $(1\times 1)$ spots (green circle) are comparatively weak. 
The $6\sqrt{3}$ spots characteristic of the ZLG buffer-layer reconstruction are strongly suppressed after intercalation, consistent with decoupling of graphene from the SiC surface and the removal of the interfacial reconstruction \cite{Riedl2009,Riedl2010,Rosenzweig2020}.

For the Ag$_{(2)}$ sample (Fig.~\ref{fig1}(b)), it exhibits, in addition to the primary diffraction spots (graphene $(1\times 1)$ and SiC $(1\times 1)$ spots indicated by the grey and green circles respectively), two distinct sets of replica spots (highlighted by pink and blue dashed box in Fig.~\ref{fig1}(b)) arising from supercells formation. 
These supercells can be clearly seen in the top-view schematics of the SiC/Ag$_{(2)}$/graphene stacking shown in Fig.~\ref{fig1}(c), based on the structural model of Ref.~\cite{Jain2025}. 
The sixfold group of satellites surrounding the graphene $(1\times 1)$ spot (in the pink box in panel(b) and the associated enlarged view) is from a $\sim (6\times 6)$ graphene supercell commensurate with a $(5\times 5)$ Ag$_{(2)}$ supercell (guided by the pink rhombus in the lower-left quadrant of Fig.~\ref{fig1}(c)).  
The second set, highlighted by the blue circles (in the blue box), is more clearly resolved in a LEED pattern (Fig.~\ref{fig1}(d)) with higher energy (\qty{94}{eV}), which is more sensitive to the SiC. 
These spots correspond to a $\sim (5\times 5)$ supercell of the SiC substrate, which is commensurate with the $(3\sqrt{3}\times 3\sqrt{3})$ supercell of Ag$_{(2)}$, as outlined by the blue rhombus in the upper-right quadrant of Fig.~\ref{fig1}(c).

To quantify the supercell periodicities, we performed Lorentzian fits to line profiles passing through the replica spots. 
Fig.~\ref{fig1}(e) shows a fit to a profile through the graphene $(1\times 1)$ spot and a diagonally opposite pair of the sixfold satellites (profile direction indicated by the pink line in Fig.~\ref{fig1}(d)). 
The average separation of the left and right replica spots from the graphene $(1\times 1)$ reflection corresponds to $0.163$ of the graphene reciprocal-lattice vector ($\mathbf{G}_{\mathrm{gr}}$), indicating an approximately $(6\times 6)$ periodicity (more precisely, $(6.1\times 6.1)$) relative to graphene.
A corresponding line profile analysis of the second set of replicas (Fig.~\ref{fig1}(f), profile direction indicated by the blue line in the enlarged view in panel(d)) yields a separation of $0.2$ of the SiC reciprocal surface lattice vector ($\mathbf{G}_{\mathrm{SiC}}$), identifying these spots as the $(5\times 5)$ SiC grid spots \cite{Starke2009}. 
Overall, the LEED signatures agree well with those reported previously for Ag$_{(2)}$ prepared using the CHet method \cite{Jain2025}. The corresponding replica features observed in ARPES are discussed later in Sec.~\ref{subsec:replica}.  

From this LEED analysis we conclude that Ag$_{(2)}$ is rotated by \ang{30} with respect to the underlying SiC lattice and does not form a simple $(1\times 1)$ epitaxial relation as in Ag$_{(1)}$, but instead adopts the commensurate relation as described above.
Maintaining the $(1\times 1)$ epitaxy in Ag$_{(1)}$ implies tensile strain, with an Ag-Ag distance (\qty{3.10} {\angstrom}) larger than that in bulk Ag (\qty{2.88} {\angstrom}) \cite{Pham2026}. In contrast, the rotated lattice of Ag$_{(2)}$ has a higher areal Ag density and a shorter Ag-Ag bond of length \qty{\sim2.98}{\angstrom} \cite{Jain2025} to help relieve the strain \cite{Jain2025, Pham2026}.

To examine the chemical purity of the prepared sample, an XPS survey spectrum acquired with \qty{650}{eV} photon energy is shown in Fig.~\ref{fig1}(g). 
The survey is dominated by the Ag~3$d_{5/2}$ and Ag~3$d_{3/2}$ spin-orbit split peak \cite{Rosenzweig2020, Bearden1967}, together with the expected C~1$s$, Si~2$p$, and Si~2$s$ emissions \cite{Bearden1967}. 
In addition, the Ag~$MNN$ Auger features \cite{Yu2022} and the valence band signal are clearly observed. 
Importantly, the O~1$s$~\cite{Bearden1967} signal is not detectable, indicating the absence of measurable air contamination. 
Furthermore, no Pb~4$f$ emission is observed in the binding energy (BE) window of $\sim$ \qtyrange{135}{143}{eV} \cite{Matta2022}. 
Given that the photoionization cross-sections of Ag~3$d$ and Pb~4$f$ are of comparable magnitude at \qty{650}{eV}~\cite{Yeh1985}, the absence of Pb~4$f$ provides strong evidence that Pb has been effectively removed during the preparation. 
We therefore conclude the above-mentioned preparation method, starting from Pb-QFMLG, yields a chemically clean Ag$_{(2)}$-ML at the graphene/SiC interface.

A pronounced photon-energy dependence is observed in the Ag~3$d$ core-level line shape when the excitation energy is reduced from \qty{650}{eV} to \qty{450}{eV}, while keeping the sample position and all other experimental parameters unchanged. 
As shown in the inset of Fig.~\ref{fig1}(g) for the Ag~3$d_{5/2}$ region, the main peak maximum remains essentially unchanged, whereas the low-BE shoulder is strongly suppressed at \qty{450}{eV}. 
To quantify this effect, the Ag~3$d_{5/2}$ spectra acquired at \qty{650}{eV} and \qty{450}{eV} were fitted using Voigt functions \cite{Olivero1977} as shown in Fig.~\ref{fig1}(h) and \ref{fig1}(i), respectively; for the wide-range fit of the Ag~3$d$ spectra, see SI, section B, Fig. S2.  
By comparing with the literature, we assign the component at \qty{\sim 368.4}{eV} to an Ag$_{(1)}$-like contribution previously reported for the Ag$_{(1)}$ phase \cite{Jain2025}. 
The dominant spectral weight is attributed to Ag$_{(2)}$$^{0}$, i.e., the in-plane \ce{Ag-Ag} bonding within the Ag$_{(2)}$ intercalation layer, while the highest-BE component is assigned to the \ce{Ag-Si} bonding at the interface with the SiC substrate \cite{Jain2025}. 
The presence of two distinct Ag$_{(2)}$ bonding components indicates a nonuniform local electronic environment within the Ag$_{(2)}$ layer, in contrast to the more homogeneous Ag$_{(1)}$ phase, which is characterized by a single Ag~3$d$ component \cite{Rosenzweig2020}.
We note a systematic BE offset of the Ag$_{(2)}$ peaks of approximately $\sim$\qty{0.3}{eV} relative to Ref.~\cite{Jain2025}, which may arise due to slight variations in the local chemical environment and screening. 
Notably, in the more surface-sensitive \qty{450}{eV} data, the relative intensity of the peak associated with the Ag$_{(1)}$ layer as compared to the Ag$_{(2)}$ components is lower than for 650 eV. 
This trend is consistent with the simplified model of the exponential attenuation of photoemission intensity with larger depth \cite{Seah1979}. Since transmission electron microscopy \cite{Jain2025} data indicates that Ag$_{(1)}$ resides deeper below the graphene overlayer than Ag$_{(2)}$, the Ag$_{(1)}$ contribution is more strongly suppressed under surface-sensitive conditions, leading to a lower relative Ag$_{(1)}$ intensity.  
As a result, the Ag$_{(1)}$ contribution is more strongly suppressed in the surface-sensitive measurement, leading to a reduced Ag$_{(1)}$/Ag$_{(2)}$ intensity ratio.
The lowest-BE component (BL-Ag) lies close to the reported value for metallic Ag \cite{Rosenzweig2020}. 
One possible origin would be non-intercalated Ag residing on top of the graphene; however, such a surface contribution would be expected to increase markedly in the more surface-sensitive \qty{450}{eV} measurement, which is not observed. 
We therefore ascribe this low-BE component to a minor phase (or domains) of intercalated Ag with a more metallic local environment.
This assignment is discussed further in connection with the last section (Sec.~\ref{subsec:Ag2_BL}).

\subsection{ARPES of Ag$_{(1)}$ and Ag$_{(2)}$} 
\begin{figure*}[htbp] 
	\vspace{-0.5 cm}
	\centering
	\includegraphics[width=1.05\textwidth]{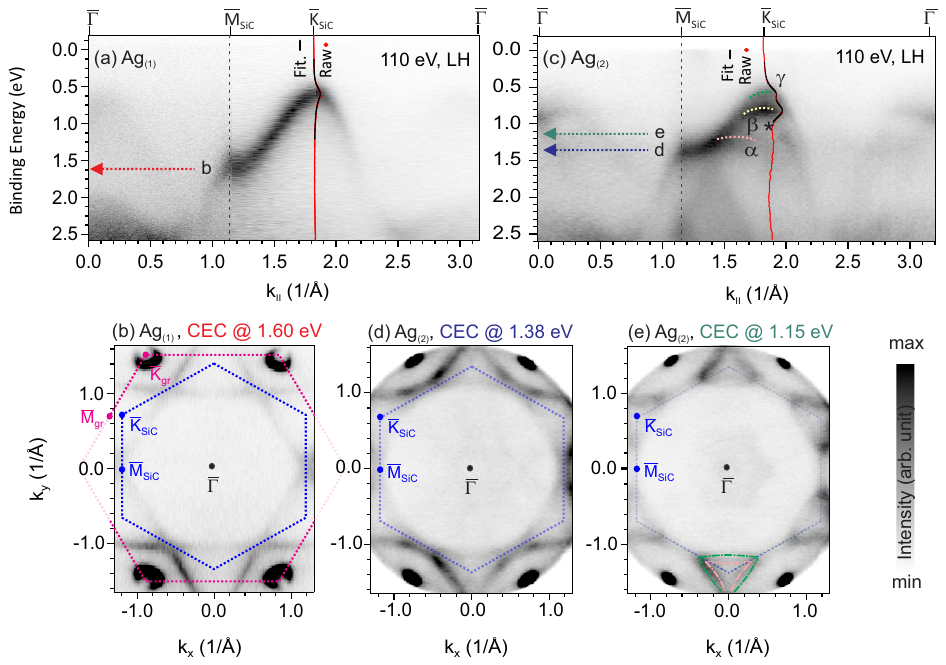}
	\vspace{-16.7 cm}
	\caption{
		$E$-$k$ dispersions of (a) Ag$_{(1)}$-QFMLG and (c) Ag$_{(2)}$-QFMLG measured along the $\overline{\Gamma\mathrm{M_{SiC}}\mathrm{K_{SiC}}\Gamma}$ direction using 110~eV photons (with linear horizontal (LH) light, i.e., p-polarization). 
		EDCs at $\overline{\mathrm{K}}_\mathrm{SiC}$ are overlaid in each panel (raw data: red points; Lorentzian fit: black line). The $\alpha$, $\beta$, and $\gamma$ band splitting, and the $\ast$-marked opening are discussed in the main text.
		(b) CEC of Ag$_{(1)}$-QFMLG at a BE of \qty{1.60}{eV}; the BZs of SiC (blue) and graphene (pink) are indicated. 
		(d,e) CEC of Ag$_{(2)}$-QFMLG at BEs of \qty{1.38}{eV} and \qty{1.15}{eV}, respectively; The CEC BEs are indicated in panels (a) and (c); two distinct Ag$_{(2)}$ pockets are highlighted in (e).  }
	\label{fig2}
\end{figure*}
Fig.~\ref{fig2}(a) presents the electronic band structure of the Ag$_{(1)}$ phase measured along the $\overline{\Gamma\mathrm{M_{SiC}}\mathrm{K_{SiC}}\Gamma}$ direction. 
The data were acquired with a photon energy of \qty{110}{eV}. 
The corresponding surface Brillouin zones (BZs) of SiC and graphene are superimposed in the constant-energy contour (CEC) data in Fig.~\ref{fig2}(b). 
Consistent with earlier reports (Ref.~\cite{Rosenzweig2020}), the Ag$_{(1)}$ phase exhibits semiconducting behavior \cite{Lee2022} as the valence band maximum (VBM) is at a BE of $0.59 \pm 0.05$~eV below the Fermi level ($E_F$) at the $\overline{\mathrm{K}}_\mathrm{SiC}$ point. 
That exact value of the VBM is confirmed by the fitting of the energy-distribution curve (EDC) superimposed at $\overline{\mathrm{K}}_\mathrm{SiC}$. 
The dispersion also displays a saddle point feature at the $\overline{\mathrm{M}}_\mathrm{SiC}$ point at a BE of approximately \qty{1.60}{eV}. 
Along $\overline{\Gamma\mathrm{M}}_\mathrm{SiC}$, Ag-derived states show clear hybridization with SiC bulk bands, which become visible near \qty{\sim 2.5}{eV} BE. 
The CEC at \qty{1.60}{eV} BE in Fig.~\ref{fig2}(b) highlights the band topology at the saddle point $\overline{\mathrm{M}}_\mathrm{SiC}$, where the contour shows a pronounced hexagonal shape. 

Fig.~\ref{fig2}(c) shows the electronic band structure of the Ag$_{(2)}$ phase measured with \qty{110} {eV} light. 
Similar to the Ag$_{(1)}$ phase, Ag$_{(2)}$ exhibits semiconducting character, with the VBM located well below $E_F$ at the $\overline{\mathrm{K}}_\mathrm{SiC}$ point. 
In contrast to Ag$_{(1)}$, however, the Ag$_{(2)}$-derived band near $\overline{\mathrm{K}}_\mathrm{SiC}$ splits into three components, as demonstrated by the EDC extracted at $\overline{\mathrm{K}}_\mathrm{SiC}$ and overlaid at the corresponding momentum. 
From this EDC fitting, the VBM is found at a BE of \qty{0.57\pm0.05}{eV}  (associated with the $\gamma$ branch), while the component with the highest spectral weight appears at \qty{0.82\pm0.05}{eV} (the $\beta$ branch). 
Along $\overline{\mathrm{M}_{\mathrm{SiC}}\mathrm{K}}_{\mathrm{SiC}}$ direction, the lowest-BE component is labeled as $\alpha$. 
A gap-like suppression of spectral weight is observed between the $\beta$ branch and the downward-dispersing branch along $\overline{\mathrm{K}_{\mathrm{SiC}}\Gamma}$ at the position marked by `$\ast$'.
A saddle-point feature is observed at the $\overline{\mathrm{M}}_\mathrm{SiC}$ point at a BE of approximately \qty{1.40}{eV}. 
The CEC in Fig.~\ref{fig2}(d) shows that, at the saddle point energy, the Ag$_{(2)}$ pockets exhibit a much weaker hexagonal warping as compared to Ag$_{(1)}$. 
A more detailed inspection also reveals two distinct Ag-derived pockets around $\overline{\mathrm{K}}_{\mathrm{SiC}}$ at a slightly higher BE, which are clearly resolved in the CEC at \qty{1.15}{eV} BE in Fig.~\ref{fig2}(e): the pocket outlined by the green dash-dotted line corresponds to the $\gamma$ band, while the inner pocket marked in pink is associated with the $\alpha$ branch.

\subsection{The periodicity in the $k$-space of Ag$_{(2)}$}
\begin{figure*}[htbp]
	\centering
	\includegraphics[width=\textwidth]{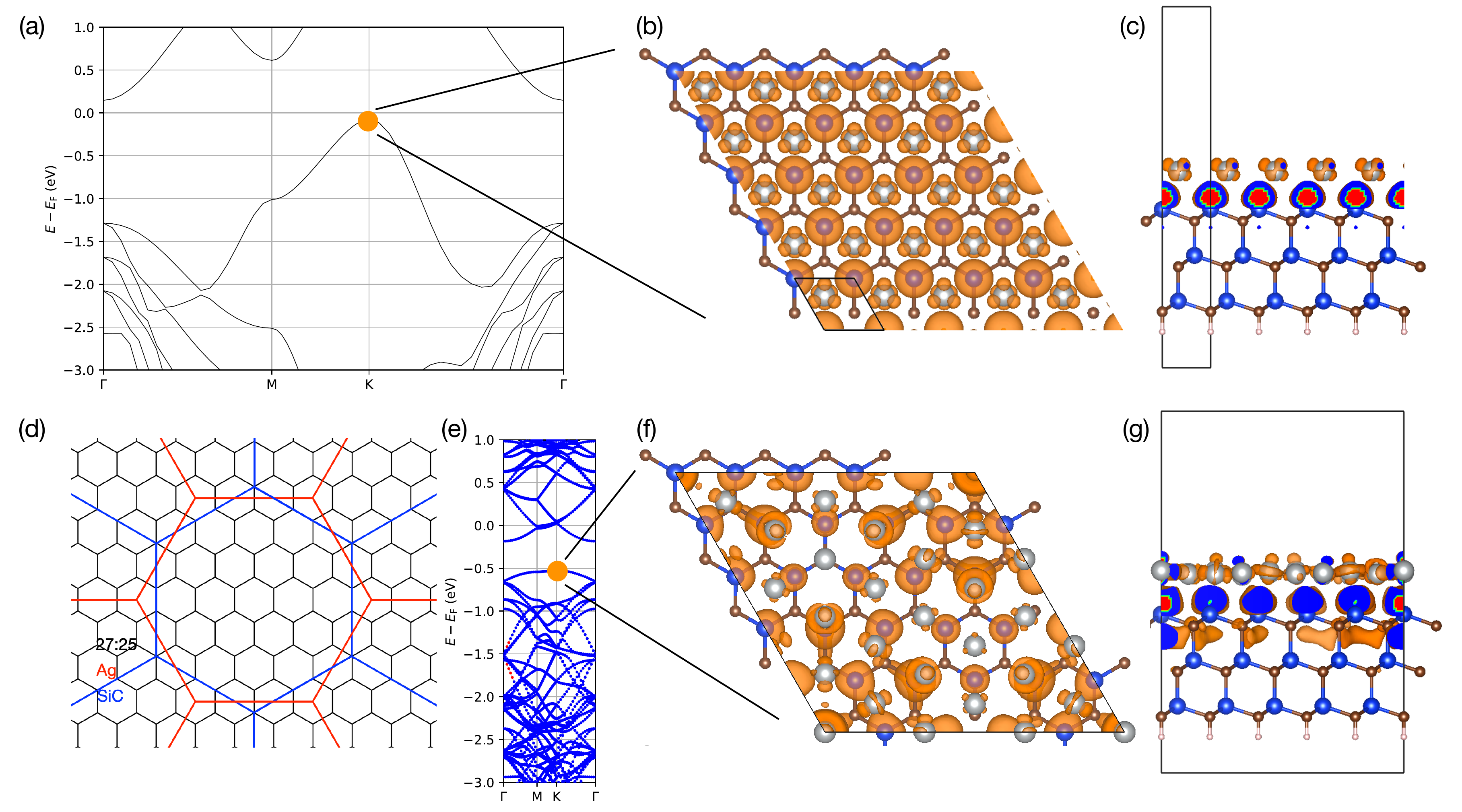}  
	\caption{
		(a) The band structure for the \ce{Ag_{(1)}} phase (with Ag atoms registered at the hollow site of the topmost SiC since it is the configuration with the lowest energy).
		(b) Top and (c) side view of the partial charge density of the state at the VBM (the orange dot) in (a).
		The orange color shows the shape of the isosurface while the red-to-blue color in the cross section shows the charge density distribution from high to low.
		(d) The Brillouin zones of the supercell for \ce{Ag$_{(2)}$} (black, marked as ``27:25'' for the Ag:SiC ratio), and the primitive Brillouin zones of the SiC (blue) and Ag (red) lattice.
		(e) The band structure for the supercell of the \ce{Ag$_{(2)}$} phase.
		The blue/red dots indicate that the corresponding state is more suitable to be unfolded to the SiC/Ag primitive cell, respectively, with a relatively smaller unfolding entropy as defined in Eq.~\eqref{eq:unfold_entropy}. 
		An example of a red dot would comparatively have more density in the Ag layer, and that part follows the periodicity of Ag (SI, section C, Fig.~S3).
		(f) Top and (g) side view of the partial charge density of the state (color code as in panels (b,c)) at the VBM (the orange dot) in (e).
	}
	\label{fig:unfolding}
\end{figure*}

Although the Ag$_{(2)}$ pockets are more circular overall, the momentum locations of the VBM and the saddle point remain the same as those of Ag$_{(1)}$.
More importantly, the periodicity in the $k$-space follows the BZ of the SiC, suggesting a $(1\times 1)$ epitaxial relationship of Ag$_{(2)}$ with the SiC substrate.
However, the structural model from Fig.~\ref{fig1}(c) shows that the Ag unit cell in Ag$_{(2)}$ is \ang{30} rotated with respect to the SiC unit cell, thus, the periodicity in the $k$-space is expected to follow the BZ indicated for the Ag (red) in Fig.~\ref{fig:unfolding}(d).
The apparent $(1\times 1)$-like symmetry in the ARPES data can be explained by the Ag-SiC interaction through inspecting the wavefunctions of the Ag$_{(2)}$ phase.
For example, we can plot and compare the partial charge density at the VBM of the Ag$_{(1)}$ phase (Fig.~\ref{fig:unfolding}(a--c)) and the Ag$_{(2)}$ phase (Fig.~\ref{fig:unfolding}(e--g)).
Although the wavefunction in Ag$_{(2)}$ has some charge density in the Ag layer, following the periodicity of the Ag lattice, the wavefunction is predominantly localized on the Si atoms, showing great similarity with that of the \ce{Ag_{(1)}} phase, therefore making the wavefunction more compatible with the SiC periodicity, meaning the VBM in Ag$_{(2)}$ should be better unfolded to the SiC primitive cell.

Computationally, to compare the calculated band structure of the Ag$_{(2)}$ phase (Fig.~\ref{fig:unfolding}(e)) with the experimental ARPES data, the bands are first calculated in the BZ of the supercell (black in Fig.~\ref{fig:unfolding}(d)), i.e., $(3\sqrt3\times3\sqrt3)$-Ag on a $(5\times5)$-SiC, and then unfolded to bands to a primitive cell \cite{popescuExtractingEffectiveBand2012}.
Here, we have two options: the SiC primitive cell and the Ag primitive cell, and we need to decide which one is more suitable for the unfolding.
For each electronic state $\vb{K}m$ in the supercell, the procedure of unfolding it to $\vb{k_j}m$ in the primitive cell (Eq.~(14, 15) in Ref. \cite{popescuExtractingEffectiveBand2012}) is essentially to redistribute the spectral weight
\begin{equation}
	1 = \sum_{\vb{G}} \ab|C_{\vb{k}m}(\vb{G})|^2
\end{equation}
from the supercell first Brillouin zone (FBZ) to the primitive FBZ
\begin{gather}
	P_{\vb{K}m}(\vb{k}_j = \vb{K} + \vb{G}_j) = \sum_{\vb{g}} \ab|C_{\vb{K}m}(\vb{g}+\vb{G}_j)|^2 \\
	\sum_j P_{\vb{K}m}(\vb{k}_j) = 1
\end{gather}
where $m$ is the band index, $C$'s are Fourier coefficients, $\vb{K}$ is a wavevector in the supercell FBZ, $\vb{G}$'s are reciprocal lattice vectors of the supercell, and their non-capital counterparts are for the primitive cell.
For an ideal supercell created simply by multiplying a primitive cell, the unfolding of a non-degenerate state would yield a single $\vb{k}_j$ with $P_{\vb{K}m}(\vb{k}_j) = 1$.
If we choose another primitive cell not compatible with the original one, we would have the spectral weight distributed over multiple $\vb{k}_j$'s, and the unfolding would be less meaningful.
To quantify how well a state is compatible with a given primitive cell, we can define an unfolding entropy as
\begin{equation}\label{eq:unfold_entropy}
	S_{\vb{K}m} = -\sum_j P_{\vb{K}m}(\vb{k}_j) \ln P_{\vb{K}m}(\vb{k}_j)
\end{equation}
with a smaller value indicating a more meaningful unfolding, and the wavefunction more compatible with the chosen primitive cell, while a value of zero indicating a perfect unfolding.
Applying this procedure to the Ag$_{(2)}$ supercell band structure (Fig.~\ref{fig:unfolding}(e)), we find that most of the states near the VBM have a smaller unfolding entropy for the SiC primitive cell (blue dots) compared to the Ag primitive cell (red dots), indicating that for these states, they are more compatible with the periodicity of SiC, and the unfolding to the SiC primitive cell is more suitable for comparison with ARPES. 
The possible reason for this might be that the high energy dangling bonds of the topmost Si are more likely to capture most of the bonding electrons when interacting with the 2D Ag.
This could be seen that even at those red dots, we still have a significant contribution from the topmost Si (See SI, section C, Fig.~S3).

\subsection{Bands comparison between the theory and the experiment}
\begin{figure*}[htbp]
	\vspace{-0.1 cm}
	\centering
	\includegraphics[width=1\textwidth]{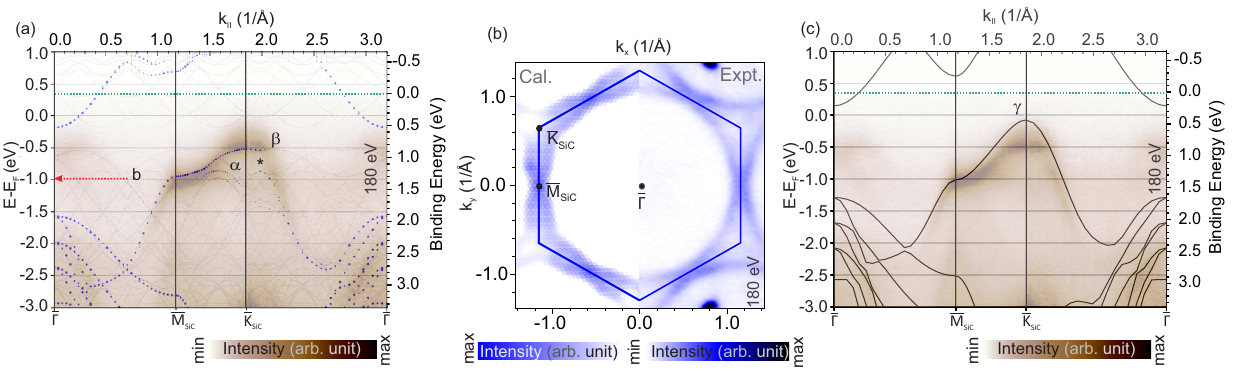}
	\vspace{-20.3 cm}
	\caption{Comparison of DFT and ARPES data.
		(a) DFT band structure of Ag$_{(2)}$ unfolded to SiC primitive cell (blue points) together with the corresponding folded supercell bands (transparent gray) along the $\overline{\Gamma\mathrm{M_{SiC}}\mathrm{K_{SiC}}\Gamma}$ direction; the ARPES spectrum measured at \qty{180}{eV} is overlaid (in brown scale) for comparison.
		(b) Comparison of (left) the calculated CEC of the unfolded Ag$_{(2)}$ bands at \qty{0.05}{eV} below the saddle point at $\overline{\mathrm{M}}_\mathrm{SiC}$ and (right) the experimental CEC at a BE of \qty{1.35} {eV}. 
		(c) DFT band structure of Ag$_{(1)}$ on SiC shown in the same format as in Fig.~\ref{fig:unfolding}(a), stacked with the Ag$_{(2)}$ ARPES data along the $\overline{\Gamma\mathrm{M_{SiC}}\mathrm{K_{SiC}}\Gamma}$ direction; The experimental Fermi level is marked by the green dashed line in (a) and (c). The left and right $y$ axes in (a) and (c) correspond to the calculated and experimental energy scales, respectively.}
	\label{fig4}
\end{figure*}

We now turn to a detailed comparison between the experimental and theoretically calculated band dispersions of the Ag$_{(2)}$ phase in Fig.~\ref{fig4}(a) (the corresponding analysis for Ag$_{(1)}$ is presented in SI, section D and Fig.~S4).
The calculated and unfolded bands are plotted along the $\overline{\Gamma\mathrm{M_{SiC}}\mathrm{K_{SiC}}\Gamma}$ direction.
Experimental data in the same direction was measured at \qty{180}{eV} and shows essentially the same dispersion as the \qty{110}{eV} data in Fig.~\ref{fig2}(c) but has higher intensity.
Overall, the DFT results reproduce the key experimental features with good agreement, including the dominant weakly dispersive (flat) $\beta$ band and the downward-dispersing $\alpha$ branch in the vicinity of $\overline{\mathrm{K}}_{\mathrm{SiC}}$.
We note that experimental intensities can vary depending  on the momentum position of the measurement, e.g., first or repeated BZ (see SI, section E, Fig. S5).
As mentioned in the Methods, we ignore the conduction bands and the position of the Fermi level in the calculated bands.
By comparing with the unfolded DFT band structure (Fig.~\ref{fig4}(a)), the ``opening'' (marked by `$\ast$' in the figure) is not really the breaking of a single band, it is rather the intensity change of two bands in the supercell:  
Along the direction from $\overline{\mathrm{K}}_\mathrm{{SiC}}$ to $\overline{\Gamma}$, the $\beta$ band loses intensity at the ``opening'', while another band (downward branch) starts to show strong intensity there.
This intensity suppression is a good example showing that the bands in ARPES are strongly influenced by the matrix element effects. 
In the calculation, the matrix element effect is partially captured by ``absorbing the additional structure factor $\braket<kn|Km>$'' between the supercell and primitive cell wavefunctions \cite{Ku2010} through unfolding the supercell band structure properly to the SiC primitive cell.  
In Fig.~\ref{fig4}(b), we compare the experimental CEC map at \qty{1.35}{eV} BE with the CEC calculated from the unfolded Ag$_{(2)}$ DFT band structure at \qty{0.05}{eV} below the saddle point at $\overline{\mathrm{M}}_\mathrm{{SiC}}$.
The calculated contour reproduces the nearly circular Ag$_{(2)}$ pocket shape in good agreement with the experiment. 
To make the CEC better match the experimental data, the DFT CEC is broadened using a Lorentzian energy smearing of \qty{0.1}{eV} and a Gaussian smearing of \qty{0.06}{\per\angstrom} along the momentum axes, followed by adjusting the intensity threshold. 
The calculated CEC with no smearing in the momentum plane and no intensity truncation is shown in Fig.~S6 (d). 

The uppermost $\gamma$ branch at $\overline{\mathrm{K}}_{\mathrm{SiC}}$ and its continuation along $\overline{\mathrm{K_{SiC}}\Gamma}$ (better seen in Fig.~\ref{fig2}(c)) are not captured by the Ag$_{(2)}$-only calculation.
Notably, these additional experimental features closely resemble the DFT band structure of the Ag$_{(1)}$ phase (see Fig.~\ref{fig4}(c)). 
SI section G and Fig.~S8 further analyzes the Ag$_{(1)}$-like contribution observed in the experimental Ag$_{(2)}$ CEC.
These observations suggest a residual mixture of the two phases in the sample, consistent with the phase coexistence inferred from the Ag~3$d$ line-shape analysis in XPS. 
Nevertheless, the relative spectral weight of the Ag-derived states indicates that Ag$_{(2)}$ remains the predominant phase in the probed region.




\subsection{Dirac cone dispersion of Ag$_{(1)}$ and Ag$_{(2)}$-QFMLG}
\begin{figure*}[htbp]
	\vspace{-0.6 cm}
	\centering
	\includegraphics[width=1\textwidth]{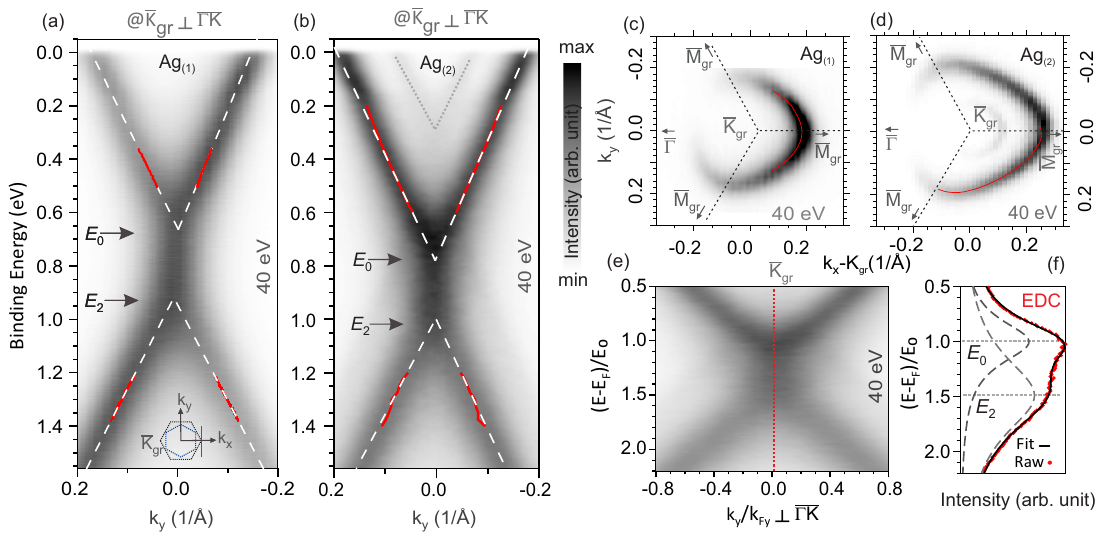}
	\vspace{-17.2 cm}
	\caption{
	(a) and (b) $E$-$k$ cuts through the graphene Dirac cone at $\overline{\mathrm{K}}_{\mathrm{gr}}$ for (a) Ag$_{(1)}$-QFMLG and (b) Ag$_{(2)}$-QFMLG, measured with photon energy 40~eV. 
	MDC fits (red markers) and their linear extrapolations (white dashed lines) are used to determine the Dirac-points $E_0$ and $E_2$. The measurement direction is indicated in the inset Brillouin zone map. 
	(c) and (d) Fermi pockets associated with the Dirac cones of (c) Ag$_{(1)}$-QFMLG and (d) Ag$_{(2)}$-QFMLG; radial MDC fits are shown by the red curves. 
	(e) Normalized (reduced-axis) representation of the Ag$_{(2)}$-QFMLG Dirac-cone dispersion for analyzing the plasmaron-related ``diamond'' feature. 
	(f) EDC (red dotted line) at $\overline{\mathrm{K}}_{\mathrm{gr}}$ with Lorentzian fit (black line); the fitted components (gray dashed line) identify the bare and plasmaronic Dirac-points, $E_0$ and $E_2$, respectively for Ag$_{(2)}$-QFMLG.  }
	\label{fig5}
\end{figure*}
The denser lattice arrangement of the Ag$_{(2)}$ phase compared to Ag$_{(1)}$ not only produces a distinct electronic structure within the Ag monolayer itself, but is also expected to modify the electronic properties of the overlying graphene layer via proximity effects \cite{Rosenzweig_overdoping, matta_charge_2025}. 
To examine this, Fig.~\ref{fig5} compares the $\pi$-band dispersion of the QFMLG layer for samples intercalated in the Ag$_{(1)}$ and Ag$_{(2)}$ phases. Figures~\ref{fig5}(a,b) show energy--momentum cuts through the graphene Dirac cone for Ag$_{(1)}$-QFMLG and Ag$_{(2)}$-QFMLG, respectively. 
The spectra were recorded at the graphene $\overline{\mathrm{K}}_{\mathrm{gr}}$ point along a momentum direction perpendicular to $\overline{\Gamma\mathrm{K}}_{\mathrm{gr}}$ using 40~eV photons. 
In both cases, the Dirac cone exhibits a characteristic renormalization near the Dirac point, manifested as an elongated crossing also known as a ``diamond-like'' dispersion \cite{Bostwick2010}. 
This behavior is consistent with electron-plasmon coupling in graphene, which gives rise to plasmaron quasiparticles and produces two apparent branches: the ``bare'' Dirac dispersion (upper branch) and an interaction-renormalized (plasmaronic) dispersion with a Dirac-point shifted to lower BE \cite{Bostwick2010, Walter2011}. 
To extract the relevant energy scales, the dispersions of the upper $\pi^\ast$ and lower $\pi$ bands were fitted independently using momentum-distribution curves (MDCs); the resulting fits are shown by the red lines. 
Linear extrapolations of these fitted dispersions (white dashed lines) were then used to determine the Dirac-point energies associated with the bare Dirac cone, $E_0$, and the plasmaronic branch, $E_2$. For Ag$_{(1)}$-QFMLG [Fig.~\ref{fig5}(a)], the bare Dirac-point energy is $E_0 = 0.66 \pm 0.02$~eV below $E_F$, whereas for Ag$_{(2)}$-QFMLG [Fig.~\ref{fig5}(b)] it shifts to $E_0 = 0.77 \pm 0.02$~eV, indicating stronger $n$-type doping in the Ag$_{(2)}$-intercalated sample.

A quantitative estimate of the carrier concentration was obtained from the Fermi surface contours shown in Fig.~\ref{fig5}(c) (Ag$_{(1)}$-QFMLG) and Fig.~\ref{fig5}(d) (Ag$_{(2)}$-QFMLG), which correspond to the $\pi^\ast$ band pocket around $\overline{\mathrm{K}}_{\mathrm{gr}}$. Radial MDC fitting of the pocket was performed (red curve) to determine the enclosed area $A$.
Using Luttinger's theorem \cite{Luttinger1960, LuttingerPhysRev.119.1153}, the electron density was evaluated from the Fermi surface area (for a spin- and valley-degenerate Dirac cone) as
\begin{equation}
	n = \frac{g_s g_v\,A}{(2\pi)^2} = \frac{A}{\pi^2},
\end{equation}
where $g_s=2$ and $g_v=2$. 
This analysis yields $n \approx 3.38\times10^{13}$~cm$^{-2}$ for Ag$_{(1)}$-QFMLG and $n \approx 5.92\times10^{13}$~cm$^{-2}$ for Ag$_{(2)}$-QFMLG, confirming the higher electron doping in the Ag$_{(2)}$-intercalated graphene. 
The enhanced $n$-type doping for Ag$_{(2)}$-QFMLG is presumably caused by a more efficient charge transfer from the intercalated Ag layer to graphene. 
In particular, the reduced graphene-Ag$_{(2)}$ separation (as suggested by the XPS analysis and previously reported TEM results \cite{Jain2025}) would strengthen the interfacial interaction and increase the net electron transfer into the graphene $\pi^\ast$ states, thereby shifting the Dirac point to higher BE. 
Note that the grey dashed cone in Fig.~\ref{fig5}(b) and the extra faint inner pocket in Fig.~\ref{fig5}(d) corresponds to monolayer graphene from the overgrown area of the sample.

We now analyze the pronounced elongation of the Dirac-point region in Ag$_{(2)}$-QFMLG, which we attribute to the intertwining of the plasmaronic Dirac cone (lower branch) with the bare Dirac cone (upper branch) of graphene. 
From the scaled energy separation between the Dirac points of the bare and plasmaronic bands, $\delta E=\left|\frac{E_2-E_0}{E_0}\right|$, we determine the effective coupling constant, $\alpha_G$ of the graphene layer, following the one-particle Green's-function analysis within the $G_0W$-RPA framework \cite{ Walter2011}. 
In simple terms, a larger $\alpha_G$ corresponds to stronger electron-plasmon coupling in graphene. 
The normalized Dirac-cone dispersion of Ag$_{(2)}$-QFMLG is shown in Fig.~\ref{fig5}(e). 
The vertical axis is plotted as the scaled energy ($\delta E$) and the horizontal axis corresponds to the reduced Fermi momentum $k_y/k_{F,y}$. Here $k_{F,y}$ denotes the $k_y$ value (at $k_x=\mathrm{K_{gr}}$) where the $\pi$ band crosses the Fermi level along the cut perpendicular to the $\overline{\Gamma\mathrm{K}}_{\mathrm{gr}}$ direction. For improved visibility of this crossing, Fig.~\ref{fig5}(e) is shown over the \qtyrange{0.5}{2.3}{eV} energy window.
For the scaling we use the bare Dirac point energy $E_0 = 0.68 \pm 0.02$~eV obtained from an EDC analysis at $\overline{\mathrm{K}}_{\mathrm{gr}}$ [Fig.~\ref{fig5}(b) \cite{Note_E0Shift}], and $k_{F,y}=0.18$~\AA$^{-1}$. 
From the resulting plasmaron ``diamond,'' we extract $\delta E = 0.49 \pm 0.04$, which corresponds to an effective coupling constant of $\alpha_G \approx 0.5$ for Ag$_{(2)}$-QFMLG. 
This value is larger than that obtained for Ag$_{(1)}$-QFMLG \cite{Rosenzweig2022}, indicating stronger electron-plasmon coupling in the Ag$_{(2)}$-intercalated system. 
Within this picture, the enhanced coupling is associated with a reduced dielectric screening provided by the Ag$_{(2)}$/SiC environment. 
Using $\alpha_G$, the effective dielectric constant of the Ag$_{(2)}$/SiC medium can be estimated as $\epsilon_{\mathrm{Ag_{(2)}}} \approx \frac{4.4}{\alpha_G}-1$, which yields $\epsilon_{\mathrm{Ag_{(2)}}} \approx 7.8$. 
This value is smaller than the corresponding estimate \cite{Note_sameAg1} for Ag$_{(1)}$-QFMLG, $\epsilon_{\mathrm{Ag_{(1)}}} \approx 10$, as shown in Ref.\cite{Rosenzweig2022}. 
The reduced $\epsilon_{\mathrm{Ag_{(2)}}}$ suggests weaker overall effective screening at the interface, which can be attributed to a more effective passivation of polarizable Si dangling-bond \cite{Bostwick2010} states after Ag$_{(2)}$ intercalation on SiC compared to the Ag$_{(1)}$/SiC interface.

\subsection{Signature of band replicas in the Ag$_{(2)}$ intercalated QFMLG}
\label{subsec:replica}
\begin{figure}
	\vspace{-0.7cm}
	\centering
	\includegraphics[width=1.1\textwidth]{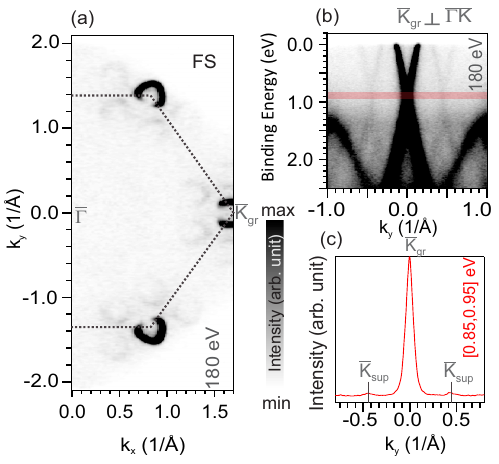}
	\vspace{-20.2 cm}
	\caption{
	(a) Fermi surface map of Ag$_{(2)}$-QFMLG measured at 180~eV, highlighting replica pockets surrounding the main graphene Dirac-cone pocket. 
	(b) $E$-$k$ cut at $\overline{\mathrm{K}}_{\mathrm{gr}}$ along a momentum direction perpendicular to $\overline{\Gamma \mathrm{K}}_{\mathrm{gr}}$, showing Dirac-cone replicas on both sides of the primary dispersion. 
	(c) MDC integrated over the energy window indicated in (b), resolving the intense main peak of the Dirac cone at $\overline{\mathrm{K}}_{\mathrm{gr}}$ and the replica features at $\overline{\mathrm{K}}_{\mathrm{sup}}$.
    }
	\label{fig6}
\end{figure}
The supercell periodicity in the Ag$_{(2)}$-QFMLG, is also evident in replicated Dirac cone pockets, as shown in Fig.~\ref{fig6}(a) by a wide-range Fermi surface map revealing such replicas surrounding the main graphene $\pi^\ast$ pocket. 
This map is measured with 180~eV photon energy. 
The replica pockets are displaced relative to the main pocket along the graphene reciprocal-lattice directions by a small momentum separation, indicative of Umklapp scattering of the graphene $\pi$-bands by the superlattice reciprocal vector, $\mathbf{G}_{\mathrm{sup}}$. 
To quantify the momentum separation, Fig.~\ref{fig6}(b) shows a two-dimensional $E$-$k$ cut extracted from Fig.~\ref{fig6}(a) through $\overline{\mathrm{K}}_{\mathrm{gr}}$ along a momentum direction perpendicular to $\overline{\Gamma \mathrm{K}}_{\mathrm{gr}}$. 
From the MDC in Fig.~\ref{fig6}(c), obtained by integrating over the \qtyrange{0.85}{0.95}{eV} BE window, we extract a supercell reciprocal-lattice vector magnitude of $|\mathbf{G}_{\mathrm{sup}}|\approx 0.49$~\AA$^{-1}$. 
This implies a real-space supercell lattice constant of approximately six times that of graphene, consistent with the $(6\times 6)$ coincidence supercell between Ag$_{(2)}$ and graphene inferred from the LEED data. 
Notably, no such replica pockets are observed for the Ag$_{(1)}$-intercalated sample. 

\subsection{Spatial dependence of Ag$_{(2)}$}
\label{subsec:Ag2_BL}
\begin{figure}
	\vspace{-0.5 cm}
	\centering
	\includegraphics[width=1.05\textwidth]{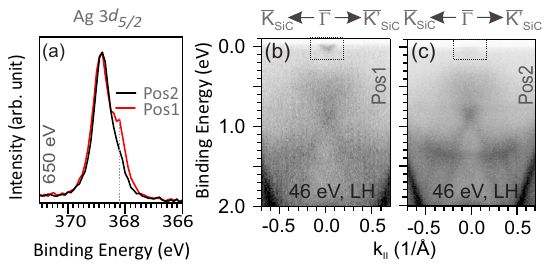}
	\vspace{-22.8 cm}
	\caption{
	(a) Ag~3$d_{5/2}$ XPS spectra of Ag$_{(2)}$-QFMLG measured at 650~eV at two sample locations (Pos1 and Pos2). 
	(b,c) ARPES spectra measured using 46~eV light (LH polarized) along the $\overline{{\mathrm{K}}_{\mathrm{SiC}}\Gamma{\mathrm{K'}}}_{\mathrm{SiC}}$ direction at (b) Pos1 and (c) Pos2, highlighting the additional band crossing $E_F$ at $\overline{\Gamma}$ in (b) that is absent in (c).}
	\label{fig7}
\end{figure}

Finally, we address the origin of the lowest-BE shoulder observed in the Ag~3$d$ core-level spectrum of the Ag$_{(2)}$-intercalated sample [Fig.~\ref{fig1}(g-i)]. As argued above, this component likely originates from a distinct, spatially localized region within the intercalated Ag layer. 
To further test this interpretation, Fig.~\ref{fig7}(a) compares the Ag~3$d_{5/2}$ spectra measured at two different sample positions (Pos1 and Pos2) using the more surface-sensitive \qty{450}{eV} excitation. 
A pronounced reduction of the shoulder intensity (BE position is marked by the dashed line) is observed when moving from Pos1 (the initial measurement position) to Pos2. 
We further compare ARPES spectra at the same positions using \qty{46}{eV} photon energy, as shown in Fig.~\ref{fig7}(b) and Fig.~\ref{fig7}(c), respectively. 
At Pos1, an additional band crossing the $E_F$ is clearly observed at the $\Gamma$ point, whereas this feature is absent at Pos2.  
A closely related behavior has been reported for Au intercalation at the graphene/SiC interface, where intercalated monolayer Au yields a semiconducting dispersion, while intercalated bilayer Au produces an additional metallic band crossing $E_F$\cite{Forti2020}. 
By analogy, we tentatively assign both the low-BE Ag~3$d$ shoulder and the $\Gamma$-centered metallic band to regions containing a bilayer (or locally thicker) intercalated Ag phase. This assignment is also consistent with the magnitude of the core-level shift: the extra Ag~3$d$ component is displaced by nearly \qty{\sim1}{eV} to lower BE relative to the main Ag$_{(2)}$~3$d$ peak, comparable to the low-BE shift reported for Au. 
Note that in Fig.~\ref{fig7}(b) and (c), the band near \qty{1}{eV} BE at $\overline{\Gamma}$ corresponds to Ag replica bands, as discussed in detail in SI Section.~F (Fig.~S6 and S7).
A question might arise whether such bilayer-like regions should also produce an observable change in the graphene doping level like Au-QFMLG. 
Here it is important to note that the spectral weight of the ``bilayer'' component is small: at Pos1 its intensity is below \qty{15}{\percent} of the total Ag$_{(2)}$~3$d$ intensity and appears confined to a highly localized region (of order \qty{\sim10}{\um}) where both the strongest $\Gamma$-crossing band and the enhanced low-BE shoulder are detected. 
At other locations--representative of most of the sample--this fraction is even smaller, as indicated by Fig.~\ref{fig7}(a) and Fig.~\ref{fig7}(c). It is therefore likely that the majority of the ARPES data discussed above were acquired in regions where the bilayer contribution is negligible, consistent with the observation of a single Dirac cone characteristic of Ag$_{(2)}$-QFMLG with predominantly monolayer Ag intercalation. 
Enhancing and controlling the bilayer-like intercalated Ag contribution, and exploring the associated electronic properties, remains an open topic for future work.

\section{Conclusion}
 
In summary, we show that changing the starting substrate from ZLG to Pb-QFMLG enables the formation of two distinct phases of intercalated 2D-Ag at the graphene/SiC interface under UHV conditions, and we provide a detailed structural and electronic property study of the second phase. 
LEED demonstrates that the Ag$_{(2)}$ layer is rotated by \ang{30} with respect to the SiC substrate and forms supercells with both SiC ($(5\times 5)$ periodicity) and graphene ($(6\times 6)$ periodicity). 
Despite the rotated Ag$_{(2)}$ lattice, high-resolution ARPES reveals a semiconducting Ag$_{(2)}$ band structure with the VBM at $\overline{\mathrm{K}}_\mathrm{SiC}$ and a saddle point near $\overline{\mathrm{M}}_\mathrm{SiC}$, i.e., at momentum positions similar to those of the Ag$_{(1)}$ phase. 
The apparent similarity of the periodicity in the $k$-space between Ag$_{(1)}$ and Ag$_{(2)}$ is rationalized by DFT: By defining the unfolding entropy, we find that, in a quantified way, for Ag$_{(2)}$, the SiC unit cell, instead of the Ag unit cell, is more suitable for most of the electronic states near VBM to be unfolded to.
This means that those electronic states have a major part with the periodicity compatible to the SiC substrate, therefore making the unfolded bands have the Ag$_{(1)}$ (or SiC) periodicity in the $k$-space. 
The calculated band structure based on the Ag$_{(2)}$ structural model is in good agreement with the experimentally observed dispersion. 
Importantly, we also find that these two Ag phases tune the electronic properties of the overlying graphene layer differently. 
Ag$_{(2)}$ produces stronger $n$-type doping in QFMLG and enhances the plasmaron-related renormalization of the Dirac point, reflected in an increase of the effective coupling constant from $\alpha_G \!\approx\! 0.4$ (Ag$_{(1)}$) to $\alpha_G \!\approx\! 0.5$ (Ag$_{(2)}$), consistent with a smaller effective dielectric constant of the Ag$_{(2)}$/SiC layers. 
In addition, ARPES resolves $(6\times 6)$ Dirac-cone replica pockets for Ag$_{(2)}$-QFMLG, providing a more precise quantitative confirmation of the graphene/Ag$_{(2)}$ supercell inferred from LEED.

More broadly, our results demonstrate that modifying the defect density of the initial substrate can stabilize different intercalated phases under UHV conditions. This provides a practical pathway to access and compare multiple 2D phases for other intercalants as well. 
Since the graphene doping level and many-body renormalizations depend strongly on the specific intercalated phase, controlling the phase of the confined layer is not only essential for fundamental studies but also critical for applications that rely on reproducible electronic tuning.

\section{Acknowledgements}
S.D.\, K.K.\, and U.St.\ acknowledge support from the Deutsche Forschungsgemeinschaft (DFG, German Research Foundation) within Research Unit FOR5242 (Project Sta315/13-1 and Ku4228/1-1).
B.Z.\ and V.H.C.\ thank the support from Two-Dimensional Crystal Consortium-Materials Innovation Platform (2DCC-MIP) under NSF cooperative agreement no.\ DMR-2039351. 
A.J. and J.A.R. acknowledge the support from National Science Foundation (NSF) award no. DMR-2011839, through the Penn State MRSEC–Center for Nanoscale Science. 
We acknowledge MAX~IV Laboratory for beamtime at the BLOCH beamline under Proposal No.~20241262, and Helmholtz-Zentrum Berlin for beamtime at the UE112-PGM-2a-$1^2$ beamline under Proposal No.~Photons-231-12042. 
We thank Dr.~Nisha Ranjan and Ms.~Vibha Reddy (MPI-FKF), Dr.~Jacek Osiecki (BLOCH beamline), Dr.~Andrei Varykhalov (UE112-PGM-2a-$1^2$ beamline), and the beamline staff at BLOCH and UE112-PGM-2a-$1^2$ for their assistance during the beamtimes. 
We also thank Ms.~Bharti Matta (MPI-FKF) for helpful discussions related to the experimental data analysis.
A.J.\ and J.A.R.\ would like to acknowledge Li-Syuan Lu, Chengye Dong and Shengxi Huang for fruitful discussions during the experimental design.
Research conducted at MAX IV, a Swedish national user facility, is supported by the Swedish Research
council under contract 2018-07152, the Swedish Governmental Agency for Innovation Systems under contract 2018-04969, and Formas under contract 2019-02496.

\bibliography{Ag2}

@article{Novoselov2004,
	author  = {Novoselov, K. S. and Geim, A. K. and Morozov, S. V. and Jiang, D. and Zhang, Y. and Dubonos, S. V. and Grigorieva, I. V. and Firsov, A. A.},
	title   = {Electric Field Effect in Atomically Thin Carbon Films},
	journal = {Science},
	volume  = {306},
	number  = {5696},
	pages   = {666--669},
	year    = {2004},
	doi     = {10.1126/science.1102896}
}

@article{Geim2007,
	author  = {Geim, A. K. and Novoselov, K. S.},
	title   = {The rise of graphene},
	journal = {Nat. Mater.},
	volume  = {6},
	number  = {3},
	pages   = {183--191},
	year    = {2007},
	doi     = {10.1038/nmat1849}
}

@article{Son2006,
	title = {Energy Gaps in Graphene Nanoribbons},
	author = {Son, Young-Woo and Cohen, Marvin L. and Louie, Steven G.},
	journal = {Phys. Rev. Lett.},
	volume = {97},
	issue = {21},
	pages = {216803},
	numpages = {4},
	year = {2006},
	month = {Nov},
	publisher = {American Physical Society},
	doi = {10.1103/PhysRevLett.97.216803},
	url = {https://link.aps.org/doi/10.1103/PhysRevLett.97.216803}
}

@article{Bhimanapati2015,
	author = {Bhimanapati, Ganesh R. and Lin, Zhong and Meunier, Vincent and Jung, Yeonwoong and Cha, Judy and Das, Saptarshi and Xiao, Di and Son, Youngwoo and Strano, Michael S. and Cooper, Valentino R. and Liang, Liangbo and Louie, Steven G. and Ringe, Emilie and Zhou, Wu and Kim, Steve S. and Naik, Rajesh R. and Sumpter, Bobby G. and Terrones, Humberto and Xia, Fengnian and Wang, Yeliang and Zhu, Jun and Akinwande, Deji and Alem, Nasim and Schuller, Jon A. and Schaak, Raymond E. and Terrones, Mauricio and Robinson, Joshua A.},
	title = {Recent Advances in Two-Dimensional Materials beyond Graphene},
	journal = {ACS Nano},
	volume = {9},
	number = {12},
	pages = {11509-11539},
	year = {2015},
	doi = {10.1021/acsnano.5b05556},
	URL = {https://doi.org/10.1021/acsnano.5b05556}	
}

@article{Lin2016,
	doi = {10.1088/2053-1583/3/4/042001},
	url = {https://doi.org/10.1088/2053-1583/3/4/042001},
	year = {2016},
	month = {dec},
	publisher = {IOP Publishing},
	volume = {3},
	number = {4},
	pages = {042001},
	author = {Lin, Zhong and McCreary, Amber and Briggs, Natalie and Subramanian, Shruti and Zhang, Kehao and Sun, Yifan and Li, Xufan and Borys, Nicholas J and Yuan, Hongtao and Fullerton-Shirey, Susan K and Chernikov, Alexey and Zhao, Hui and McDonnell, Stephen and Lindenberg, Aaron M and Xiao, Kai and LeRoy, Brian J and Drndić, Marija and Hwang, James C M and Park, Jiwoong and Chhowalla, Manish and Schaak, Raymond E and Javey, Ali and Hersam, Mark C and Robinson, Joshua and Terrones, Mauricio},
	title = {2D materials advances: from large scale synthesis and controlled heterostructures to improved characterization techniques, defects and applications},
	journal = {2D Mater.},
}

@article{Bera2010,
	author  = {Bera, Debasis and Qian, Ling and Tseng, Tze and Holloway, Paul H.},
	title   = {Quantum dots and their multimodal applications: a review},
	journal = {Materials (Basel)},
	volume  = {3},
	number  = {4},
	pages   = {2260-2345},
	year    = {2010},
	doi     = {doi:10.3390/ma3042260}
}

@article{Huang2022,
	title = {{2D} semiconductors for specific electronic applications: from device to system},
	volume = {6},
	issn = {2397-7132},
	url = {https://doi.org/10.1038/s41699-022-00327-3},
	doi = {10.1038/s41699-022-00327-3},
	number = {1},
	journal = {npj 2D Mater. Appl.},
	author = {Huang, Xiaohe and Liu, Chunsen and Zhou, Peng},
	month = aug,
	year = {2022},
	pages = {51},
}

@article{Novoselov2005,
	author  = {Novoselov, K. S. and Geim, A. K. and Morozov, S. V. and Jiang, D. and Katsnelson, M. I. and Grigorieva, I. V. and Dubonos, S. V. and Firsov, A. A.},
	title   = {Two-dimensional gas of massless Dirac fermions in graphene},
	journal = {Proc. Natl. Acad. Sci.},
	volume  = {102},
	number  = {30},
	pages   = {10451--10453},
	year    = {2005},
	doi     = {10.1073/pnas.0502848102}
}

@article{karakachian_one-dimensional_2020,
	title = {One-dimensional confinement and width-dependent bandgap formation in epitaxial graphene nanoribbons},
	volume = {11},
	issn = {2041-1723},
	url = {https://doi.org/10.1038/s41467-020-19051-x},
	doi = {10.1038/s41467-020-19051-x},
	number = {1},
	journal = {Nat. Commun.},
	author = {Karakachian, Hrag and Nguyen, T. T. Nhung and Aprojanz, Johannes and Zakharov, Alexei A. and Yakimova, Rositsa and Rosenzweig, Philipp and Polley, Craig M. and Balasubramanian, Thiagarajan and Tegenkamp, Christoph and Power, Stephen R. and Starke, Ulrich},
	month = dec,
	year = {2020},
	pages = {6380},
}

@article{Parzinger2017,
	author  = {Parzinger, Eric and Mitterreiter, Elmar and Stelzer, Max and Kreupl, Franz and Ager, Joel W. and Holleitner, Alexander W. and Wurstbauer, Ursula},
	title   = {Hydrogen evolution activity of individual mono-, bi-, and few-layer {MoS}$_2$ towards photocatalysis},
	journal = {Appl. Mater. Today},
	volume  = {8},
	pages   = {132--140},
	year    = {2017},
	doi     = {10.1016/j.apmt.2017.04.007}
}

@article{Lee2008,
	author  = {Lee, Changgu and Wei, Xiaoding and Kysar, Jeffrey W. and Hone, James},
	title   = {Measurement of the Elastic Properties and Intrinsic Strength of Monolayer Graphene},
	journal = {Science},
	volume  = {321},
	number  = {5887},
	pages   = {385--388},
	year    = {2008},
	doi     = {10.1126/science.1157996}
}

@article{Mak2010,
	title = {Atomically Thin ${\mathrm{MoS}}_{2}$: A New Direct-Gap Semiconductor},
	author = {Mak, Kin Fai and Lee, Changgu and Hone, James and Shan, Jie and Heinz, Tony F.},
	journal = {Phys. Rev. Lett.},
	volume = {105},
	issue = {13},
	pages = {136805},
	numpages = {4},
	year = {2010},
	month = {Sep},
	publisher = {American Physical Society},
	doi = {10.1103/PhysRevLett.105.136805},
	url = {https://link.aps.org/doi/10.1103/PhysRevLett.105.136805}
}

@article{Bostwick2010,
	author  = {Bostwick, Aaron and Speck, Florian and Seyller, Thomas and Horn, Karsten and Rotenberg, Eli},
	title   = {Observation of Plasmarons in Quasi-Freestanding Doped Graphene},
	journal = {Science},
	volume  = {328},
	number  = {5981},
	pages   = {999--1002},
	year    = {2010},
	doi     = {10.1126/science.1186489}
}

@article{Rosenzweig2022,
	title = {Surface charge-transfer doping a quantum-confined silver monolayer beneath epitaxial graphene},
	author = {Rosenzweig, Philipp and Karakachian, Hrag and Marchenko, Dmitry and Starke, Ulrich},
	journal = {Phys. Rev. B},
	volume = {105},
	issue = {23},
	pages = {235428},
	numpages = {12},
	year = {2022},
	month = {Jun},
	publisher = {American Physical Society},
	doi = {10.1103/PhysRevB.105.235428},
	url = {https://link.aps.org/doi/10.1103/PhysRevB.105.235428}
}

@article{Link2019,
	title = {Introducing strong correlation effects into graphene by gadolinium intercalation},
	author = {Link, S. and Forti, S. and St\"ohr, A. and K\"uster, K. and R\"osner, M. and Hirschmeier, D. and Chen, C. and Avila, J. and Asensio, M. C. and Zakharov, A. A. and Wehling, T. O. and Lichtenstein, A. I. and Katsnelson, M. I. and Starke, U.},
	journal = {Phys. Rev. B},
	volume = {100},
	issue = {12},
	pages = {121407},
	numpages = {6},
	year = {2019},
	month = {Sep},
	publisher = {American Physical Society},
	doi = {10.1103/PhysRevB.100.121407},
	url = {https://link.aps.org/doi/10.1103/PhysRevB.100.121407}
}

@article{Herrera2024,
	author  = {Herrera, Sa{\'u}l A. and Parra-Mart{\'i}nez, Guillermo and Rosenzweig, Philipp and Matta, B. and Polley, C. M. and K{\"u}ster, K. and Starke, Ulrich and Guinea, Francisco and Silva-Guill{\'e}n, J. {\'A}. and Naumis, Gerardo G. and Pantale{\'o}n, P. A.},
	title   = {Topological Superconductivity in Heavily Doped Single-Layer Graphene},
	journal = {ACS Nano},
	volume  = {18},
	number  = {51},
	pages   = {34842--34857},
	year    = {2024},
	doi     = {10.1021/acsnano.4c12532}
}

@article{Zhong2017,
	author  = {Zhong, Q. and Zhang, J and Cheng, P and Feng, B and Li, W and Sheng, S and Li,  H and Meng, S and Chen, L and Wu, K.},
	title   = {Metastable phases of 2D boron sheets on Ag(111)},
	journal = {J. Phys.: Condens. Matter},
	volume  = {29},
	pages   = {095002},
	year    = {2017},
	doi     = {10.1088/1361-648X/aa5165}
}

@article{Romanyuk2009,
	title = {Nanoscale Pit Formation at 2D Ge Layers on Si: Influence of Energy and Entropy},
	author = {Romanyuk, Konstantin and Brona, Jacek and Voigtl\"ander, Bert},
	journal = {Phys. Rev. Lett.},
	volume = {103},
	issue = {9},
	pages = {096101},
	numpages = {4},
	year = {2009},
	month = {Aug},
	publisher = {American Physical Society},
	doi = {10.1103/PhysRevLett.103.096101},
	url = {https://link.aps.org/doi/10.1103/PhysRevLett.103.096101}
}

@article{Koerner2011,
	title = {Second-Layer Induced Island Morphologies in Thin-Film Growth of Fullerenes},
	author = {K\"orner, Martin and Loske, Felix and Einax, Mario and K\"uhnle, Angelika and Reichling, Michael and Maass, Philipp},
	journal = {Phys. Rev. Lett.},
	volume = {107},
	issue = {1},
	pages = {016101},
	numpages = {4},
	year = {2011},
	month = {Jun},
	publisher = {American Physical Society},
	doi = {10.1103/PhysRevLett.107.016101},
	url = {https://link.aps.org/doi/10.1103/PhysRevLett.107.016101}
}

@article{Aizawa2014,
	author  = {Aizawa, Takashi and Suehara, Shigeru and Otani, Shigeki},
	title   = {Silicene on Zirconium Carbide (111)},
	journal = {J. Phys. Chem. C},
	volume  = {118},
	number  = {40},
	pages   = {23049--23057},
	year    = {2014},
	doi     = {10.1021/jp505602c}
}

@article{Riedl2009,
	title = {Quasi-Free-Standing Epitaxial Graphene on SiC Obtained by Hydrogen Intercalation},
	author = {Riedl, C. and Coletti, C. and Iwasaki, T. and Zakharov, A. A. and Starke, U.},
	journal = {Phys. Rev. Lett.},
	volume = {103},
	issue = {24},
	pages = {246804},
	numpages = {4},
	year = {2009},
	month = {Dec},
	publisher = {American Physical Society},
	doi = {10.1103/PhysRevLett.103.246804},
	url = {https://link.aps.org/doi/10.1103/PhysRevLett.103.246804}
}

@article{Emtsev2011,
	title = {Ambipolar doping in quasifree epitaxial graphene on SiC(0001) controlled by Ge intercalation},
	author = {Emtsev, Konstantin V. and Zakharov, Alexei A. and Coletti, Camilla and Forti, Stiven and Starke, Ulrich},
	journal = {Phys. Rev. B},
	volume = {84},
	issue = {12},
	pages = {125423},
	numpages = {6},
	year = {2011},
	month = {Sep},
	publisher = {American Physical Society},
	doi = {10.1103/PhysRevB.84.125423},
	url = {https://link.aps.org/doi/10.1103/PhysRevB.84.125423}
}

@article{Forti2020,
	author  = {Forti, Stiven and Link, Stefan and St{\"o}hr, Alexander and Niu, Yuran and Zakharov, Alexei A. and Coletti, Camilla and Starke, Ulrich},
	title   = {Semiconductor to metal transition in two-dimensional gold and its van der Waals heterostack with graphene},
	journal = {Nat. Commun.},
	volume  = {11},
	pages   = {2236},
	year    = {2020},
	doi     = {10.1038/s41467-020-15683-1}
}

@article{Rosenzweig2020,
	title = {Large-area synthesis of a semiconducting silver monolayer via intercalation of epitaxial graphene},
	author = {Rosenzweig, Philipp and Starke, Ulrich},
	journal = {Phys. Rev. B},
	volume = {101},
	issue = {20},
	pages = {201407},
	numpages = {6},
	year = {2020},
	month = {May},
	publisher = {American Physical Society},
	doi = {10.1103/PhysRevB.101.201407},
	url = {https://link.aps.org/doi/10.1103/PhysRevB.101.201407}
}

@article{Matta2022,
	title = {Momentum microscopy of Pb-intercalated graphene on SiC: Charge neutrality and electronic structure of interfacial Pb},
	author = {Matta, Bharti and Rosenzweig, Philipp and Bolkenbaas, Olaf and K\"uster, Kathrin and Starke, Ulrich},
	journal = {Phys. Rev. Res.},
	volume = {4},
	issue = {2},
	pages = {023250},
	numpages = {11},
	year = {2022},
	month = {Jun},
	publisher = {American Physical Society},
	doi = {10.1103/PhysRevResearch.4.023250},
	url = {https://link.aps.org/doi/10.1103/PhysRevResearch.4.023250}
}

@article{albalushi2016,
	title = {Two-dimensional gallium nitride realized via graphene encapsulation},
	volume = {15},
	issn = {1476-4660},
	url = {https://doi.org/10.1038/nmat4742},
	doi = {10.1038/nmat4742},
	number = {11},
	journal = {Nat. Mater.},
	author = {Al Balushi, Zakaria Y. and Wang, Ke and Ghosh, Ram Krishna and Vilá, Rafael A. and Eichfeld, Sarah M. and Caldwell, Joshua D. and Qin, Xiaoye and Lin, Yu-Chuan and DeSario, Paul A. and Stone, Greg and Subramanian, Shruti and Paul, Dennis F. and Wallace, Robert M. and Datta, Suman and Redwing, Joan M. and Robinson, Joshua A.},
	month = nov,
	year = {2016},
	pages = {1166--1171},
}

@article{El-Sherif2021,
	author = {El-Sherif, Hesham and Briggs, Natalie and Bersch, Brian and Pan, Minghao and Hamidinejad, Mahdi and Rajabpour, Siavash and Filleter, Tobin and Kim, Ki Wook and Robinson, Joshua and Bassim, Nabil D.},
	title = {Scalable Characterization of 2D Gallium-Intercalated Epitaxial Graphene},
	journal = {ACS Appl. Mater. Interfaces},
	volume = {13},
	number = {46},
	pages = {55428-55439},
	year = {2021},
	doi = {10.1021/acsami.1c14091},
	URL = {https://doi.org/10.1021/acsami.1c14091}
}

@article{Vera2024,
	author = {Vera, Alexander and Zheng, Boyang and Yanez, Wilson and Yang, Kaijie and Kim, Seong Yeoul and Wang, Xinglu and Kotsakidis, Jimmy C. and El-Sherif, Hesham and Krishnan, Gopi and Koch, Roland J. and Bowen, T. Andrew and Dong, Chengye and Wang, Yuanxi and Wetherington, Maxwell and Rotenberg, Eli and Bassim, Nabil and Friedman, Adam L. and Wallace, Robert M. and Liu, Chaoxing and Samarth, Nitin and Crespi, Vincent H. and Robinson, Joshua A.},
	title = {Large-Area Intercalated Two-Dimensional Pb/Graphene Heterostructure as a Platform for Generating Spin–Orbit Torque},
	journal = {ACS Nano},
	volume = {18},
	number = {33},
	pages = {21985-21997},
	year = {2024},
	doi = {10.1021/acsnano.4c04075},
	URL = {https://doi.org/10.1021/acsnano.4c04075}	
}

@article{Wundrack2026,
	author = {Wundrack, Stefan and Bothe, Marc and Jaime, Marcelo and Küster, Kathrin and Gruschwitz, Markus and Yin, Yefei and Mamiyev, Zamin and Schädlich, Philip and Matta, Bharti and Datta, Sawani and Eckert, Marius and Tegenkamp, Christoph and Starke, Ulrich and Stosch, Rainer and Schumacher, Hans Werner and Seyller, Thomas and Pierz, Klaus and Tschirner, Teresa and Bakin, Andrey},
	title = {Lithographically Controlled Liquid Metal Diffusion in Graphene: Fabrication and Magnetotransport Signatures of Superconductivity},
	journal = {Adv. Mater.},
	volume = {38},
	number = {5},
	pages = {e11992},
	doi = {https://doi.org/10.1002/adma.202511992},
	url = {https://advanced.onlinelibrary.wiley.com/doi/abs/10.1002/adma.202511992},
	year = {2026}
}

@article{Briggs2020,
	title = {Atomically thin half-van der {Waals} metals enabled by confinement heteroepitaxy},
	volume = {19},
	issn = {1476-4660},
	url = {https://doi.org/10.1038/s41563-020-0631-x},
	doi = {10.1038/s41563-020-0631-x},
	number = {6},
	journal = {Nat. Mater.},
	author = {Briggs, Natalie and Bersch, Brian and Wang, Yuanxi and Jiang, Jue and Koch, Roland J. and Nayir, Nadire and Wang, Ke and Kolmer, Marek and Ko, Wonhee and De La Fuente Duran, Ana and Subramanian, Shruti and Dong, Chengye and Shallenberger, Jeffrey and Fu, Mingming and Zou, Qiang and Chuang, Ya-Wen and Gai, Zheng and Li, An-Ping and Bostwick, Aaron and Jozwiak, Chris and Chang, Cui-Zu and Rotenberg, Eli and Zhu, Jun and van Duin, Adri C. T. and Crespi, Vincent and Robinson, Joshua A.},
	month = jun,
	year = {2020},
	pages = {637--643},
}

@article{Wetherington2021,
	author  = {Wetherington, Maxwell T. and Turker, Furkan and Bowen, Timothy and Vera, Alexander and Rajabpour, Siavash and Briggs, Natalie and Subramanian, Shruti and Maloney, Ashley and Robinson, Joshua A.},
	title   = {2-dimensional polar metals: a low-frequency Raman scattering study},
	journal = {2D Mater.},
	volume  = {8},
	number  = {4},
	pages   = {041003},
	year    = {2021},
	doi     = {10.1088/2053-1583/ac2245}
}

@article{Zhang2025,
	author = {Kunyan Zhang  and Rinu Abraham Maniyara  and Yuanxi Wang  and Arpit Jain  and Maxwell T. Wetherington  and Thuc T. Mai  and Chengye Dong  and Timothy Bowen  and Ke Wang  and Slava V. Rotkin  and Angela R. Hight Walker  and Vincent H. Crespi  and Joshua Robinson  and Shengxi Huang },
	title = {Tunable phononic quantum interference induced by two-dimensional metals},
	journal = {Sci. Adv.},
	volume = {11},
	number = {32},
	pages = {eadw1800},
	year = {2025},
	doi = {10.1126/sciadv.adw1800},
	URL = {https://www.science.org/doi/abs/10.1126/sciadv.adw1800}
}

@article{Liu2025,
	author = {Liu, Matthew W.-J. and Ulman, Kanchan Ajit and Zheng, Boyang and Jain, Arpit and Heintzelman, Daniel J. and Wang, Ke and He, Wen and Dong, Chengye and Lu, Li-Syuan and Crespi, Vincent H. and Quek, Su Ying and Robinson, Joshua A. and Knappenberger, Kenneth L. Jr.},
	title = {Structure-Dependent Electronic Relaxation Dynamics of Two-Dimensional Silver Monolayers},
	journal = {Nano Letters},
	volume = {25},
	number = {49},
	pages = {17145-17151},
	year = {2025},
	doi = {10.1021/acs.nanolett.5c04723},
	URL = {https://doi.org/10.1021/acs.nanolett.5c04723},
}

@article{Jain2025,
	author        = {Jain, Arpit and Zheng, Boyang and Datta, Sawani and Ulman, Kanchan and Henz, Jakob and Liu, Matthew Wei-Jun and Pham, Van Dong and He, Wen and Dong, Chengye and Lu, Li-Syuan and Vera, Alexander and Sawtarie, Nader and Auker, Wesley and Wang, Ke and Hengstebeck, Bob and Henshaw, Zachary W. and Mathela, Shreya and Wetherington, Maxwell T. and Blades, William H. and Knappenberger, Kenneth and Wurstbauer, Ursula and Quek, Su Ying and Starke, Ulrich and Huang, Shengxi and Crespi, Vincent H. and Robinson, Joshua A.},
	title         = {Defect-Mediated Phase Engineering of 2D Ag at the Graphene/SiC Interface},
	journal       = {arXiv},
	year          = {2025},
	eprint        = {2511.07151},
	archivePrefix = {arXiv},
	primaryClass  = {cond-mat.mtrl-sci},
	url           = {https://arxiv.org/abs/2511.07151}
}

@article{Dong2024,
	author = {Dong, Chengye and Lu, Li-Syuan and Lin, Yu-Chuan and Robinson, Joshua A.},
	title = {Air-Stable, Large-Area 2D Metals and Semiconductors},
	journal = {ACS Nanoscience Au},
	volume = {4},
	number = {2},
	pages = {115-127},
	year = {2024},
	doi = {10.1021/acsnanoscienceau.3c00047},	
	URL = {https://doi.org/10.1021/acsnanoscienceau.3c00047},
}

@article{Pompei2025,
	title = {Novel {Structures} of {Gallenene} {Intercalated} in {Epitaxial} {Graphene}},
	volume = {21},
	url = {https://onlinelibrary.wiley.com/doi/abs/10.1002/smll.202505640},
	doi = {https://doi.org/10.1002/smll.202505640},
	number = {38},
	journal = {Small},
	author = {Pompei, Emanuele and Skibińska, Katarzyna and Senesi, Giulio and Vlamidis, Ylea and Rossi, Antonio and Forti, Stiven and Coletti, Camilla and Beltram, Fabio and Rubini, Silvia and Sorba, Lucia and Heun, Stefan and Veronesi, Stefano},
	year = {2025},
	pages = {e05640},
}

@article{Ramachandran1998,
	author  = {Ramachandran, V. and Brady, M. F. and Smith, A. R. and Feenstra, R. M. and Greve, D. W.},
	title   = {Preparation of Atomically Flat Surfaces on Silicon Carbide Using Hydrogen Etching},
	journal = {J. Electron. Mater.},
	volume  = {27},
	pages   = {308--312},
	year    = {1998},
	doi     = {10.1007/s11664-998-0406-7}
}

@article{soubatch2005,
	author = {Soubatch, S. and Saddow, Stephen E. and Rao, Shailaja P. and Lee, W.Y. and Konuma, M. and Starke, Ulrich},
	title = {Structure and Morphology of 4H-SiC Wafer Surfaces after H2-Etching},
	year = {2005},
	month = {2},
	volume = {483},
	pages = {761--764},
	journal = {Mater. Sci. Forum},
	doi = {10.4028/www.scientific.net/MSF.483-485.761}
}

@article{Emtsev2009,
	author  = {Emtsev, K. V. and Bostwick, A. and Horn, K. and Jobst, J. and Kellogg, G. L. and Ley, L. and McChesney, J. L. and Ohta, T. and Reshanov, S. A. and R{\"o}hrl, J. and Rotenberg, E. and Schmid, A. K. and Waldmann, D. and Weber, H. B. and Seyller, Th.},
	title   = {Towards wafer-size graphene layers by atmospheric pressure graphitization of silicon carbide},
	journal = {Nat. Mater.},
	volume  = {8},
	number  = {3},
	pages   = {203--207},
	year    = {2009},
	doi     = {10.1038/nmat2382}
}

@article{Riedl2010,
	author  = {Riedl, C. and Coletti, C. and Starke, U.},
	title   = {Structural and electronic properties of epitaxial graphene on SiC(0001): a review of growth, characterization, transfer doping and hydrogen intercalation},
	journal = {J. Phys. D: Appl. Phys.},
	volume  = {43},
	number  = {37},
	pages   = {374009},
	year    = {2010},
	doi     = {10.1088/0022-3727/43/37/374009}
}

@article{Forti2014,
	doi = {10.1088/0022-3727/47/9/094013},
	url = {https://doi.org/10.1088/0022-3727/47/9/094013},
	year = {2014},
	month = {feb},
	publisher = {IOP Publishing},
	volume = {47},
	number = {9},
	pages = {094013},
	author = {Forti, S and Starke, U},
	title = {Epitaxial graphene on SiC: from carrier density engineering to quasi-free standing graphene by atomic intercalation},
	journal = {J. Phys. D: Appl. Phys.}
}

@article{Emtsev2008,
	title = {Interaction, growth, and ordering of epitaxial graphene on SiC{0001} surfaces: A comparative photoelectron spectroscopy study},
	author = {Emtsev, K. V. and Speck, F. and Seyller, Th. and Ley, L. and Riley, J. D.},
	journal = {Phys. Rev. B},
	volume = {77},
	issue = {15},
	pages = {155303},
	numpages = {10},
	year = {2008},
	month = {Apr},
	publisher = {American Physical Society},
	doi = {10.1103/PhysRevB.77.155303},
	url = {https://link.aps.org/doi/10.1103/PhysRevB.77.155303}
}

@article{Matta2025,
	title = {Pb-intercalated epitaxial graphene on SiC: Full insight into band structure and orbital character of interlayer Pb, and charge transfer into graphene},
	author = {Matta, Bharti and Rosenzweig, Philipp and K\"uster, Kathrin and Polley, Craig and Starke, Ulrich},
	journal = {Phys. Rev. B},
	volume = {111},
	issue = {15},
	pages = {155435},
	numpages = {11},
	year = {2025},
	month = {Apr},
	publisher = {American Physical Society},
	doi = {10.1103/PhysRevB.111.155435},
	url = {https://link.aps.org/doi/10.1103/PhysRevB.111.155435}
}

@article{Fiori2017,
	title = {Li-intercalated graphene on SiC(0001): An STM study},
	author = {Fiori, Sara and Murata, Yuya and Veronesi, Stefano and Rossi, Antonio and Coletti, Camilla and Heun, Stefan},
	journal = {Phys. Rev. B},
	volume = {96},
	issue = {12},
	pages = {125429},
	numpages = {8},
	year = {2017},
	month = {Sep},
	publisher = {American Physical Society},
	doi = {10.1103/PhysRevB.96.125429},
	url = {https://link.aps.org/doi/10.1103/PhysRevB.96.125429}
}

@article{Schaedlich2023,
	author  = {Sch{\"a}dlich, Philip and Ghosal, Chitran and Stettner, Monja and Matta, Bharti and Wolff, Susanne and Sch{\"o}lzel, Felix and Richter, Philipp and Hutter, Michael and Haags, Andreas and Wenzel, Sebastian and Mamiyev, Zaur and Koch, Jan and Soubatch, Sergej and Rosenzweig, Philipp and Polley, Craig and Tautz, Fred S. and Kumpf, Christian and K{\"u}ster, Kathrin and Starke, Ulrich and Seyller, Thomas and Bocquet, Fran{\c c}ois C. and Tegenkamp, Christoph},
	title   = {Domain Boundary Formation Within an Intercalated Pb Monolayer Featuring Charge-Neutral Epitaxial Graphene},
	journal = {Adv. Mater. Interfaces},
	volume  = {10},
	number  = {27},
	pages   = {2300471},
	year    = {2023},
	doi     = {10.1002/admi.202300471}
}

@article{Polley2024,
	author  = {Polley, Craig M. and Leandersson, Mats and Adell, Johan and Osiecki, Jakub and Carbone, Daniele and Ali, Khadiza and Fedderwitz, Heike and Balasubramanian, Thiagarajan},
	title   = {The Bloch Beamline at MAX IV: Micro-Spot ARPES from a Conventional, Full-Featured Beamline},
	journal = {Synchrotron Radiat. News},
	volume  = {37},
	number  = {4},
	pages   = {18--23},
	year    = {2024},
	doi     = {10.1080/08940886.2024.2391252}
}

@article{Varykhalov2018,
	author  = {Varykhalov, Andrei},
	title   = {$1^{2}$-ARPES: The ultra-high-resolution photoemission station at the U112-PGM-2a-$1^{2}$ beamline at BESSY II},
	journal = {J. Large-Scale Res. Facil.},
	volume  = {4},
	pages   = {A99},
	year    = {2018},
	doi     = {10.17815/jlsrf-4-99}
}

@article{Seah1979,
	author  = {Seah, M. P. and Dench, W. A.},
	title   = {Quantitative electron spectroscopy of surfaces: A standard data base for electron inelastic mean free paths in solids},
	journal = {Surf. Interface Anal.},
	volume  = {1},
	number  = {1},
	pages   = {2-11},
	year    = {1979},
	doi     = {10.1002/sia.740010103}
}

@article{Starke2009,
	author  = {Starke, U. and Riedl, C.},
	title   = {Epitaxial graphene on SiC(0001) and SiC(000\={1}): from surface reconstructions to carbon electronics},
	journal = {J. Phys.: Condens. Matter},
	volume  = {21},
	number  = {13},
	pages   = {134016},
	year    = {2009},
	doi     = {10.1088/0953-8984/21/13/134016}
}

@article{Pham2026,
	title = {Strain-induced reconstruction in two-dimensional silver intercalated between graphene and SiC},
	author = {Pham, Van Dong and Zheng, Boyang and Jain, Arpit and Dong, Chengye and Lu, Li-Syuan and Henshaw, Zachary W. and Blades, William H. and Robinson, Joshua A. and Crespi, Vincent H. and Trampert, Achim and Engel-Herbert, Roman},
	journal = {Phys. Rev. Mater.},
	volume = {10},
	issue = {3},
	pages = {034003},
	numpages = {10},
	year = {2026},
	month = {Mar},
	publisher = {American Physical Society},
	doi = {10.1103/pc8w-hz4t},
	url = {https://link.aps.org/doi/10.1103/pc8w-hz4t}
}

@article{Olivero1977,
	author  = {Olivero, J. J. and Longbothum, R. L.},
	title   = {Empirical fits to the Voigt line width: A brief review},
	journal = {J. Quant. Spectrosc. Radiat. Transfer},
	volume  = {17},
	number  = {2},
	pages   = {233--236},
	year    = {1977},
	doi     = {10.1016/0022-4073(77)90161-3}
}

@article{Bearden1967,
	author  = {Bearden, J. A. and Burr, A. F.},
	title   = {Reevaluation of X-Ray Atomic Energy Levels},
	journal = {Rev. Mod. Phys.},
	volume  = {39},
	pages   = {125--142},
	year    = {1967},
	doi     = {10.1103/RevModPhys.39.125}
}

@article{Yu2022,
	author  = {Yu, Jianzhong and Ye, Song and Xv, Xinling and Pan, Ling and Lin, Peixuan and Liao, Huazhen and Wang, Deping},
	title   = {Thermal-Driven Formation of Silver Clusters Inside Na/Li {FAU}Y Zeolites for Formaldehyde Detection},
	journal = {Nanomaterials},
	volume  = {12},
	number  = {18},
	pages   = {3215},
	year    = {2022},
	url     = {https://www.mdpi.com/2079-4991/12/18/3215}
}

@article{Yeh1985,
	author  = {Yeh, J. J. and Lindau, I.},
	title   = {Atomic subshell photoionization cross sections and asymmetry parameters: 1 \ensuremath{\le} Z \ensuremath{\le} 103},
	journal = {At. Data Nucl. Data Tables},
	volume  = {32},
	number  = {1},
	pages   = {1--155},
	year    = {1985},
	doi     = {10.1016/0092-640X(85)90016-6}
}

@article{popescuExtractingEffectiveBand2012,
	title = {Extracting {{E}} versus $\vec{k}$ Effective Band Structure from Supercell Calculations on Alloys and Impurities},
	author = {Popescu, Voicu and Zunger, Alex},
	year = 2012,
	month = feb,
	journal = {Phys. Rev. B},
	volume = {85},
	number = {8},
	pages = {085201},
	issn = {1098-0121, 1550-235X},
	doi = {10.1103/PhysRevB.85.085201},
	urldate = {2021-05-06},
	langid = {english}
}

@article{Ku2010,
	title = {Unfolding First-Principles Band Structures},
	author = {Ku, Wei and Berlijn, Tom and Lee, Chi-Cheng},
	journal = {Phys. Rev. Lett.},
	volume = {104},
	issue = {21},
	pages = {216401},
	numpages = {4},
	year = {2010},
	month = {May},
	publisher = {American Physical Society},
	doi = {10.1103/PhysRevLett.104.216401},
	url = {https://link.aps.org/doi/10.1103/PhysRevLett.104.216401}
}

@article{Rosenzweig_overdoping,
	title = {Overdoping Graphene beyond the van Hove Singularity},
	author = {Rosenzweig, Philipp and Karakachian, Hrag and Marchenko, Dmitry and K\"uster, Kathrin and Starke, Ulrich},
	journal = {Phys. Rev. Lett.},
	volume = {125},
	issue = {17},
	pages = {176403},
	numpages = {6},
	year = {2020},
	month = {Oct},
	publisher = {American Physical Society},
	doi = {10.1103/PhysRevLett.125.176403},
	url = {https://link.aps.org/doi/10.1103/PhysRevLett.125.176403}
}

@article{matta_charge_2025,
	title = {Charge transfer between van der {Waals} coupled metallic {2D} layers},
	volume = {17},
	url = {http://dx.doi.org/10.1039/D5NR01368B},
	doi = {10.1039/D5NR01368B},
	number = {33},
	journal = {Nanoscale},
	author = {Matta, Bharti and Rosenzweig, Philipp and Polley, Craig and Starke, Ulrich and Küster, Kathrin},
	year = {2025},
	publisher = {The Royal Society of Chemistry},
	pages = {19317--19323},
}

@article{Walter2011,
	title = {Effective screening and the plasmaron bands in graphene},
	author = {Walter, Andrew L. and Bostwick, Aaron and Jeon, Ki-Joon and Speck, Florian and Ostler, Markus and Seyller, Thomas and Moreschini, Luca and Chang, Young Jun and Polini, Marco and Asgari, Reza and MacDonald, Allan H. and Horn, Karsten and Rotenberg, Eli},
	journal = {Phys. Rev. B},
	volume = {84},
	issue = {8},
	pages = {085410},
	numpages = {8},
	year = {2011},
	month = {Aug},
	publisher = {American Physical Society},
	doi = {10.1103/PhysRevB.84.085410},
	url = {https://link.aps.org/doi/10.1103/PhysRevB.84.085410}
}

@article{Luttinger1960,
	title = {Ground-State Energy of a Many-Fermion System. II},
	author = {Luttinger, J. M. and Ward, J. C.},
	journal = {Phys. Rev.},
	volume = {118},
	issue = {5},
	pages = {1417--1427},
	numpages = {0},
	year = {1960},
	month = {Jun},
	publisher = {American Physical Society},
	doi = {10.1103/PhysRev.118.1417},
	url = {https://link.aps.org/doi/10.1103/PhysRev.118.1417}
}

@article{LuttingerPhysRev.119.1153,
	title = {Fermi Surface and Some Simple Equilibrium Properties of a System of Interacting Fermions},
	author = {Luttinger, J. M.},
	journal = {Phys. Rev.},
	volume = {119},
	issue = {4},
	pages = {1153--1163},
	numpages = {0},
	year = {1960},
	month = {Aug},
	publisher = {American Physical Society},
	doi = {10.1103/PhysRev.119.1153},
	url = {https://link.aps.org/doi/10.1103/PhysRev.119.1153}
}

@misc{Note_E0Shift,
	title        = {We note a small difference in the extracted bare {D}irac point energy depending on the analysis method: linear extrapolation of the {MDC}-derived dispersion yields ${E}_0 = 0.77$~e{V}, whereas the {EDC} analysis gives ${E}_0 = 0.68$~e{V}. {F}or the determination width of the diamond, we follow the standard procedure and normalize using the ${E}_0$ value obtained from the {EDC} analysis. {T}he discrepancy between the two estimates is likely related to the strong plasmaron-induced renormalization in the immediate vicinity of the {D}irac point, which can shift the apparent crossing in linear extrapolations. {A} similar offset of ${E}_0$ values is also observed for the {Ag}$_{(1)}$-intercalated sample.}
}

@misc{Note_sameAg1,
	title        = {For {Ag}$_{(1)}$-{QFMLG}, the {D}irac points extracted from our {MDC}-based linear extrapolation, ${E}_0$ and ${E}_2$, are consistent with the values reported in {R}ef.~\cite{Rosenzweig2022}. {H}owever, due to experimental constraints in the present measurements, the {D}irac cone features are broader, which reduces the precision of the normalized plasmaron-diamond analysis and the resulting plasmaron width. {S}ince ${E}_0$ and ${E}_2$ agree with {R}ef.~\cite{Rosenzweig2022}, we assume the corresponding plasmaron width to be unchanged and therefore adopt the same effective coupling constant $\alpha_{G}$ and dielectric constant {$\epsilon_{\mathrm{Ag}_{(1)}}$} as reported there.}
}

@article{kresseEfficiencyAbinitioTotal1996,
	title = {Efficiency of Ab-Initio Total Energy Calculations for Metals and Semiconductors Using a Plane-Wave Basis Set},
	author = {Kresse, G. and Furthm{\"u}ller, J.},
	year = 1996,
	month = jul,
	journal = {Comput. Mater. Sci.},
	volume = {6},
	number = {1},
	pages = {15--50},
	issn = {09270256},
	doi = {10.1016/0927-0256(96)00008-0},
	urldate = {2021-05-06},
	langid = {english}
}

@article{kresseEfficientIterativeSchemes1996,
	title = {Efficient Iterative Schemes for Ab Initio Total-Energy Calculations Using a Plane-Wave Basis Set},
	author = {Kresse, G. and Furthm{\"u}ller, J.},
	year = 1996,
	month = oct,
	journal = {Phys. Rev. B},
	volume = {54},
	number = {16},
	pages = {11169--11186},
	publisher = {American Physical Society},
	doi = {10.1103/PhysRevB.54.11169},
	urldate = {2023-10-18}
}

@article{kresseInitioMolecularDynamics1993,
	title = {{\emph{Ab Initio}} Molecular Dynamics for Liquid Metals},
	author = {Kresse, G. and Hafner, J.},
	year = 1993,
	month = jan,
	journal = {Phys. Rev. B},
	volume = {47},
	number = {1},
	pages = {558--561},
	issn = {0163-1829, 1095-3795},
	doi = {10.1103/PhysRevB.47.558},
	urldate = {2021-05-06},
	langid = {english}
}

@article{kresseUltrasoftPseudopotentialsProjector1999,
	title = {From Ultrasoft Pseudopotentials to the Projector Augmented-Wave Method},
	author = {Kresse, G. and Joubert, D.},
	year = 1999,
	month = jan,
	journal = {Phys. Rev. B},
	volume = {59},
	number = {3},
	pages = {1758--1775},
	issn = {0163-1829, 1095-3795},
	doi = {10.1103/PhysRevB.59.1758},
	urldate = {2021-05-06},
	langid = {english}
}

@article{blochlProjectorAugmentedwaveMethod1994,
	title = {Projector Augmented-Wave Method},
	author = {Bl{\"o}chl, P. E.},
	year = 1994,
	month = dec,
	journal = {Phys. Rev. B},
	volume = {50},
	number = {24},
	pages = {17953--17979},
	issn = {0163-1829, 1095-3795},
	doi = {10.1103/PhysRevB.50.17953},
	urldate = {2021-05-06},
	langid = {english}
}

@article{perdewGeneralizedGradientApproximation1996,
	title = {Generalized {{Gradient Approximation Made Simple}}},
	author = {Perdew, John P. and Burke, Kieron and Ernzerhof, Matthias},
	year = 1996,
	month = oct,
	journal = {Phys. Rev. Lett.},
	volume = {77},
	number = {18},
	pages = {3865--3868},
	issn = {0031-9007, 1079-7114},
	doi = {10.1103/PhysRevLett.77.3865},
	urldate = {2021-05-06},
	langid = {english}
}

@article{grimmeConsistentAccurateInitio2010,
	title = {A Consistent and Accurate {\emph{Ab Initio}} Parametrization of Density Functional Dispersion Correction ({{DFT-D}}) for the 94 Elements {{H-Pu}}},
	author = {Grimme, Stefan and Antony, Jens and Ehrlich, Stephan and Krieg, Helge},
	year = 2010,
	month = apr,
	journal = {J. Chem. Phys.},
	volume = {132},
	number = {15},
	pages = {154104},
	issn = {0021-9606, 1089-7690},
	doi = {10.1063/1.3382344},
	urldate = {2021-05-06},
	langid = {english}
}

@article{grimmeEffectDampingFunction2011,
	title = {Effect of the Damping Function in Dispersion Corrected Density Functional Theory},
	author = {Grimme, Stefan and Ehrlich, Stephan and Goerigk, Lars},
	year = 2011,
	month = may,
	journal = {J. Comput. Chem.},
	volume = {32},
	number = {7},
	pages = {1456--1465},
	issn = {01928651},
	doi = {10.1002/jcc.21759},
	urldate = {2021-05-06},
	langid = {english}
}

@misc{zhengVaspBandUnfolding2023,
	title = {{{VaspBandUnfolding}}},
	author = {Zheng, Qijing},
	year = 2026,
	url = {https://github.com/QijingZheng/VaspBandUnfolding}
}

@article{perdewDensityFunctionalTheory1985,
	title = {Density Functional Theory and the Band Gap Problem},
	author = {Perdew, John P.},
	year = 1985,
	journal = {Int. J. Quantum Chem.},
	volume = {28},
	number = {S19},
	pages = {497--523},
	issn = {1097-461X},
	doi = {10.1002/qua.560280846},
	urldate = {2025-06-27},
	langid = {english}
}

@article{Liu2021,
	author = {Liu, Yue and Liu, Xiaojie and Wang, Cai-Zhuang and Han, Yong and Evans, James W. and Lii-Rosales, Ann and Tringides, Michael C. and Thiel, Patricia A.},
	title = {Mechanism of Metal Intercalation under Graphene through Small Vacancy Defects},
	journal = {J. Phys. Chem. C},
	volume = {125},
	number = {12},
	pages = {6954-6962},
	year = {2021},
	doi = {10.1021/acs.jpcc.1c00814},
	URL = {https://doi.org/10.1021/acs.jpcc.1c00814}
}

@article{Chen2020,
	title = {Growth and stability of Pb intercalated phases under graphene on SiC},
	author = {Chen, S. and Thiel, P. A. and Conrad, E. and Tringides, M. C.},
	journal = {Phys. Rev. Mater.},
	volume = {4},
	issue = {12},
	pages = {124005},
	numpages = {10},
	year = {2020},
	month = {Dec},
	publisher = {American Physical Society},
	doi = {10.1103/PhysRevMaterials.4.124005},
	url = {https://link.aps.org/doi/10.1103/PhysRevMaterials.4.124005}
}

@article{Lee2022,
	author = {Lee, Woojoo and Wang, Yuanxi and Qin, Wei and Kim, Hyunsue and Liu, Mengke and Nunley, T. Nathan and Fang, Bin and Maniyara, Rinu and Dong, Chengye and Robinson, Joshua A. and Crespi, Vincent H. and Li, Xiaoqin and MacDonald, Allan H. and Shih, Chih-Kang},
	title = {Confined Monolayer Ag As a Large Gap 2D Semiconductor and Its Momentum Resolved Excited States},
	journal = {Nano Letters},
	volume = {22},
	number = {19},
	pages = {7841-7847},
	year = {2022},
	doi = {10.1021/acs.nanolett.2c02501},
	URL = {https://doi.org/10.1021/acs.nanolett.2c02501}
}

@article{Lu2026,
	author  = {Lu, Li-Syuan and Vera, Alexander and Turker, Furkan and Ananthanarayanan, Krishnan Mekkanamkulam and Dong, Chengye and Wetherington, Maxwell and Robinson, Joshua A.},
	title   = {Atomic-scale confinement at the graphene/{SiC} interface: A platform for novel two-dimensional nanomaterials},
	journal = {J. Vac. Sci. Technol. A},
	volume  = {44},
	number  = {3},
	pages   = {030802},
	year    = {2026},
	doi     = {10.1116/6.0005391}
}
\bibliographystyle{apsrev4-2}

\end{document}